\documentclass[a4paper,10pt]{article}
\pdfoutput=1
\usepackage{epsfig}
\usepackage{amssymb}
\usepackage{amsfonts}
\usepackage{amsmath}
\usepackage{blindtext}
\usepackage{hyperref}
\usepackage{esint}
\usepackage{euscript}
\usepackage{verbatim}
\usepackage{latexsym}
\usepackage{graphicx}
\usepackage{caption}
\usepackage{float}
\usepackage{subcaption}
\usepackage{bbm}

\usepackage{listings}

\usepackage{booktabs}
\hypersetup{
    colorlinks=true,
    linkcolor=black,
    filecolor=black,      
    urlcolor=black,
    pdfpagemode=FullScreen,
    }

\DeclareMathOperator{\csch}{csch}

\newif\ifdtup

\catcode`\@=11

\@addtoreset{equation}{section}

\def\@normalsize{\@setsize\normalsize{15pt}\xiipt\@xiipt
\abovedisplayskip 14pt plus3pt minus3pt%
\belowdisplayskip \abovedisplayskip
\abovedisplayshortskip \z@ plus3pt%
\belowdisplayshortskip 7pt plus3.5pt minus0pt}

\def\small{\@setsize\small{13.6pt}\xipt\@xipt
\abovedisplayskip 13pt plus3pt minus3pt%
\belowdisplayskip \abovedisplayskip
\abovedisplayshortskip \z@ plus3pt%
\belowdisplayshortskip 7pt plus3.5pt minus0pt
\def\@listi{\parsep 4.5pt plus 2pt minus 1pt
     \itemsep \parsep
     \topsep 9pt plus 3pt minus 3pt}}

\relax

\catcode`@=12

\catcode`\@=11

\def\section{\@startsection{section}{1}{\z@}{3.5ex plus 1ex minus
   .2ex}{2.3ex plus .2ex}{\large\bf}}

\def\SymBoxes#1#2#3#4{\newdimen\un@t \un@t#3%
\raisebox{#1}{\rule{#2\un@t}{#4}\hskip-#2\un@t
\@tempdimb\un@t \advance\@tempdimb by-#4\@tempcntb#2\relax%
\@whilenum{\@tempcntb>0}\do{
\rule{#4}{\un@t}\hskip\@tempdimb \advance\@tempcntb by\m@ne}%
\hskip-#2\un@t \rule[\un@t]{#2\un@t}{#4}%
\rule[\un@t]{#4}{#4}\hskip-#4
\rule{#4}{\un@t}}\hskip-#4}                

\begin{document}

\newcommand{\beq}{\begin{equation}}
\newcommand{\eeq}{\end{equation}}
\newcommand{\bea}{\begin{eqnarray}}
\newcommand{\eea}{\end{eqnarray}}
\newcommand{\beas}{\begin{eqnarray*}}
\newcommand{\eeas}{\end{eqnarray*}}
\newcommand{\defi}{\stackrel{\rm def}{=}}
\newcommand{\non}{\nonumber}
\newcommand{\bquo}{\begin{quote}}
\newcommand{\enqu}{\end{quote}}
\renewcommand{\(}{\begin{equation}}
\renewcommand{\)}{\end{equation}}
\def \eqn#1#2{\begin{equation}#2\label{#1}\end{equation}}

\def\e{\epsilon}
\def\IZ{{\mathbb Z}}
\def\IR{{\mathbb R}}
\def\IC{{\mathbb C}}
\def\IQ{{\mathbb Q}}
\def\de{\partial}
\def\Tr{ \hbox{\rm Tr}}
\def\H{ \hbox{\rm H}}
\def\HE{ \hbox{$\rm H^{even}$}}
\def\HO{ \hbox{$\rm H^{odd}$}}
\def\K{ \hbox{\rm K}}
\def\Im{ \hbox{\rm Im}}
\def\Ker{ \hbox{\rm Ker}}
\def\const{\hbox {\rm const.}}
\def\o{\over}
\def\im{\hbox{\rm Im}}
\def\re{\hbox{\rm Re}}
\def\bra{\langle}\def\ket{\rangle}
\def\Arg{\hbox {\rm Arg}}
\def\Re{\hbox {\rm Re}}
\def\Im{\hbox {\rm Im}}
\def\exo{\hbox {\rm exp}}
\def\diag{\hbox{\rm diag}}
\def\longvert{{\rule[-2mm]{0.1mm}{7mm}}\,}
\def\a{\alpha}
\def\dag{{}^{\dagger}}
\def\tq{{\widetilde q}}
\def\p{{}^{\prime}}
\def\W{W}
\def\N{{\cal N}}
\def\hsp{,\hspace{.7cm}}

\def\br{\nonumber}
\def\IZ{{\mathbb Z}}
\def\IR{{\mathbb R}}
\def\IC{{\mathbb C}}
\def\IQ{{\mathbb Q}}
\def\IP{{\mathbb P}}
\def \eqn#1#2{\begin{equation}#2\label{#1}\end{equation}}

\newcommand{\C}{\ensuremath{\mathbb C}}
\newcommand{\Z}{\ensuremath{\mathbb Z}}
\newcommand{\R}{\ensuremath{\mathbb R}}
\newcommand{\rp}{\ensuremath{\mathbb {RP}}}
\newcommand{\cp}{\ensuremath{\mathbb {CP}}}
\newcommand{\vac}{\ensuremath{|0\rangle}}
\newcommand{\vact}{\ensuremath{|00\rangle}                    }
\newcommand{\oc}{\ensuremath{\overline{c}}}
\newcommand{\psizero}{\psi_{0}}
\newcommand{\phizero}{\phi_{0}}
\newcommand{\hzero}{h_{0}}
\newcommand{\psiin}{\psi_{\rh}}
\newcommand{\phiin}{\phi_{\rh}}
\newcommand{\hin}{h_{\rh}}
\newcommand{\rh}{r_{h}}
\newcommand{\rb}{r_{b}}
\newcommand{\psibnd}{\psi_{0}^{b}}
\newcommand{\psibndp}{\psi_{1}^{b}}
\newcommand{\phibnd}{\phi_{0}^{b}}
\newcommand{\phibndp}{\phi_{1}^{b}}
\newcommand{\gbnd}{g_{0}^{b}}
\newcommand{\hbnd}{h_{0}^{b}}
\newcommand{\zh}{z_{h}}
\newcommand{\zb}{z_{b}}
\newcommand{\man}{\mathcal{M}}
\newcommand{\hbr}{\bar{h}}
\newcommand{\tbr}{\bar{t}}

\begin{titlepage}

\def\thefootnote{\fnsymbol{footnote}}

\begin{center}
{\large
{\bf Stretched Horizon from Conformal Field Theory}
}
\end{center}

\begin{center}
\ Suchetan Das\footnote{\texttt{suchetan1993@gmail.com }} 

\end{center}

\renewcommand{\thefootnote}{\arabic{footnote}}

\begin{center}

 {Center for High Energy Physics,\\
Indian Institute of Science, Bangalore 560012, India}\\

\end{center}
\vspace{-0.15in}
\noindent
\begin{center} {\bf Abstract} \end{center}
Recently, it has been observed that the Hartle-Hawking correlators, a signature of smooth horizon, can emerge from certain heavy excited state correlators in the (manifestly non-smooth) BTZ stretched horizon background, in the limit when the stretched horizon approaches the real horizon. In this note, we develop a framework of quantizing the CFT modular Hamiltonian, that explains the necessity of introducing a stretched horizon and the emergence of thermal features in the AdS-Rindler and (planar) BTZ backgrounds. In more detail, we quantize vacuum modular Hamiltonian on a spatial segment of $S^{1}$,
 which can be written as a particular linear combination of sl(2,$\mathbb{R}$) generators. 
Unlike radial quantization, (Euclidean) time circles emerge naturally here which can be contracted smoothly to the `fixed points'(end points of the interval) of this quantization thus providing a direct link to thermal physics. To define a Hilbert space with discrete normalizable states and to construct a Virasoro algebra with finite central extension, a natural regulator ($\epsilon$) is needed around the fixed points. Eventually, in the dual description the fixed points correspond to the horizons of AdS-Rindler patch or (planar) BTZ and the cut-off being the stretched horizon. 
We construct a (Lorentzian) highest weight representation of that Virasoro algebra where vacuum can be identified with certain boundary states on the cut-off surface. We further demonstrate that two point function in a (vacuum) descendant state of the regulated Hilbert space will reproduce thermal answer in $\epsilon \rightarrow 0$ limit which is analogous to the recent observation of emergent thermality in (planar) BTZ stretched horizon background. We also argue the thermal entropy of this quantization coincides with entanglement entropy of the subregion. Conversely, the microcanonical entropy corresponding to high energy density of states exactly reproduce the BTZ entropy. Quite remarkably, all these dominant high lying microstates are defined only at finite $\epsilon$ in the regulated Hilbert space. We expect that all our observations can be generalized to BTZ in stretched horizon background where the boundary spatial coordinate is compactified.

\vspace{1.6 cm}
\vfill

\end{titlepage}

\setcounter{footnote}{0}
\tableofcontents

\section{Introduction}

A key implication of the AdS/CFT correspondence\cite{Maldacena:1997re} is that the Hilbert space of quantum gravity in Asymptotically AdS (AAdS) is isomorphic to that of a unitary conformal field theory \cite{Aharony:1999ti}. The isomorphism is well-understood in the weak gravity regime $G_{N} \rightarrow 0$, wherein the effective field theory in AdS can be exactly reproduced from the large $N$ limit of holographic CFT\cite{Heemskerk:2009pn, El-Showk:2011yvt}. The standard way of describing the Hilbert space of a Euclidean CFT on a sphere is via radial quantization in Euclidean time. The vacuum of this quantization, which is left invariant under (global) conformal transformation, is dual to the global AdS spacetime (in $G_{N} \rightarrow 0$ limit). Similarly other states (primary, descendant and their linear combinatons) in this quantization will also have different AAdS representations. For instance, a TFD state above the Hawking-Page temperature is described by an eternal AdS blackhole \cite{Maldacena:2001kr}. 

This holographic duality is best understood for AdS$_{3}$/CFT$_{2}$, where the infinite enhancement of the two dimensional conformal algebra, simplifies the bulk dynamics of AAdS$_{3}$ \cite{Brown:1986nw, Banados:1994tn}. In fact, the local bulk dynamics enters only through the non-trivial asymptotic symmetries described by the boundary Virasoro algebra. In this space of large diffeomorphism (upto an orbifolding), the existence of BTZ blackhole \cite{Banados:1992gq} facilitates the study of quantum nature of black holes and the related puzzles. 
While the TFD above Hawking-Page phase transition describes an eternal BTZ, the high energy primaries(and their descendants) are expected to describe the BTZ microstates \cite{Fitzpatrick:2014vua, Fitzpatrick:2015zha} which gives rise to the asymptotic Cardy density of states\cite{Cardy:1986ie}. All of the above statements have been realized in radial quantization and in the Hilbert space associated to it.
In spite of all these identifications in the weak gravity limit, the understanding towards non-perturbative (or finite $N$) quantum gravity using the duality is still out of our reach. To demystify black hole information loss paradox in the paradigm of AdS/CFT, or to address the issues of thermal decaying nature of (AdS) black hole correlators in the strict semiclassical limit, one need to go beyond perturbative analysis where a proper understanding of finite $G_{N}$ physics is believed to be crucial \cite{Maldacena:2001kr}. A correct and precise understanding of AdS blackhole microstates and it's CFT description will definitely lead us to progress in this direction.

Recently, the observation \cite{Leutheusser:2021qhd, Leutheusser:2021frk} of the emergence of type III$_{1}$ von Neumann algebra constructed out of the single trace operators of boundary CFT on $\mathbb{R}\times S^{d-1}$(in Lorentzian) with respect to a TFD state \footnote{To be specific, the representation of single trace algebra acting on GNS Hilbert space} in large $N$ and above the Hawking-Page temperature, leads us to a stronger hint that the smooth horizon as well as the interior might be an emergent description in $G_{N} \rightarrow 0$ limit. This type III$_{1}$ algebraic structure only holds at infinite $N$ which agrees with the algebra of bulk quantum fields in the exterior of blackhole \footnote{By definition, algebra of observables of a local QFT (also applied to QFT in local region of curved spacetime) should always be type III$_{1}$ factor \cite{Witten:2018zxz}.}. From the bulk side, the essential signature of type III algebra lies in the continuum spectrum of field modes (quasinormal modes) which is true for both black holes and black branes \cite{Papadodimas:2012aq}\footnote{Also true for AdS-Rindler or any geometries having a horizon in any patches}. However for planar BTZ, the boundary algebra will always be type III due to infinite volume effect of thermal cylinder \cite{Furuya:2023fei}. To understand the emerging CFT type III algebra (from type I) corresponding to planar BTZ, will be one of our key motivations for this work.

Along this line of discussion, we have argued in one of our recent papers \cite{Burman:2023kko} \footnote{See also \cite{Banerjee:2024dpl} for a similar yet complimentary discussion.} that, by choosing a hard wall cut-off ($\epsilon$) \footnote{This is originally introduced as the brick wall model with a motivation to understand the origin of black hole entropy\cite{tHooft:1984kcu}} outside the horizon of the BTZ, one could obtain a discrete normal mode spectrum \cite{Solodukhin:2005qy, Krishnan:2023jqn} for free scalar fields which plays crucial role to obtain the BTZ entropy, Hawking temperature\cite{Mukohyama:1998rf} and Hartle Hawking correlator in the $\epsilon \rightarrow 0$ limit\footnote{See also \cite{Burman:2024egy} for the recent discussion on emergent interior in the same limit.}. In other words, we used Dirichlet boundary condition for bulk fields on the stretched horizon situated at the distance $\epsilon$ from the horizon which makes the algebra to be type I and at the end we took the limit $\epsilon \rightarrow 0$ to recover all the known properties of BTZ as well as the type III algebra in the presence of smooth horizon. In particular, the BTZ entropy coincides with the microcanonical entropy of those scalar fields at energy to be equal to the mass of the BTZ\cite{Krishnan:2023jqn}. Similarly, an excited state of the same energy scale constructed on top of the vacuum of the free scalar theory in this stretched horizon background, will behave like a  thermal Hartle-Hawking state in the $\epsilon \rightarrow 0$ limit. This provides an effective field theory description in a finite $G_{N}$ model\footnote{Note that, in \cite{Burman:2023kko} we use a Planckian stretched horizon which fixes $\epsilon \propto G_{N}^{2}$. Hence semiclassical limit is dubbed by $\epsilon \rightarrow 0$. However, the same analysis could be done without fixing $\epsilon$ which we will discuss in a later section.} which explicitly describes the type I to type III transition or the emergence of smooth horizon from explicit non-smoothness. If this is a reliable mechanism of the underlying algebraic transition, then it must have a dual CFT description. From the bulk side, the stretched horizon is a deep IR cut-off which is located just outside the horizon. In the traditional premise of radial quantization, such boundary condition is generically hard to find \footnote{A related example is the global quench in 2d CFT which can be understood as imposing boundary conditions on the Euclidean time-strip \cite{Calabrese:2016xau} and the dual is described by placing an IR end of the world brane beyond the horizon that cuts the eternal BTZ at some fixed radial direction \cite{Hartman:2013qma, Almheiri:2018ijj}.}. Also in radial quantization, there is no known one to one correspondence that can directly relate some boundary CFT parameters to the horizon in the dual geometry. Even if we could find some kind of boundary state that is exactly dual to stretched horizon, this would not explain the necessity of imposing such boundary condition purely from CFT point of view. In this note, our main goal is to find a CFT description that could explain some of the bulk features in the presence of stretched horizon as we observed in \cite{Burman:2023kko}. To be precise, we want to ask the following:
\begin{itemize}
    \item \textit{What is the CFT description that explains the appearance of stretched horizon in the gravity dual?}
    \item \textit{What is the CFT mechanism that elucidates type I to type III transition of algebra of operators by making a parallel to the bulk transition modeled in the presence of stretched horizon?}
    \item \textit{What is the description of the CFT states that naturally relates states in the bulk Fock space in stretched horizon background as in \cite{Burman:2023kko}? What would be the boundary Hilbert space description of that regulated Fock space?}
\end{itemize}

An intuitive guess to address these questions is to find a CFT framework that could capture the thermal features directly. In the usual setting of radial quantization, to study thermal 2d CFT on a line, one need to use conformal transformation from plane to thermal cylinder where the Euclidean time is periodic. In the dual AdS$_{3}$, this transformation generates a large diffeomorphism which maps Poincare AdS to planar BTZ. Hence to probe the blackhole background directly, the bulk clock in this frame must be synchronized to the boundary time where the CFT lives. In other words, this suggests we need to find a framework where CFT is quantized along the (Euclidean) time which is periodic and the vacuum state of the quantization is described by (planar) BTZ in the bulk side. This new frame of quantization should be orthogonal to standard radial quantization and must be determined by a new CFT Hamiltonian. This is generically a difficult problem to find a Hamiltonian having this features. However in \cite{Tada:2019rls}, Tada studied a class of sl(2,$\mathbb{R}$) deformed Hamiltonian where the quantization (Euclidean) time is periodic. Recently, in \cite{Agia:2022srj, Agia:2024wxx} a similar program has also been initiated where they quantize Rindler Hamiltonian (which is generated by Euclidean rotation) by foliating the 2D plane in constant angle slices. Both of these quantizations employ similar ideas, since the sl(2,$\mathbb{R}$) deformed Hamiltonian which Tada considered in \cite{Tada:2019rls}, is exactly (upto some choice of parameters) the (total) modular Hamiltonian of an interval in $S^{1}$ for the CFT vacuum state as observed in \cite{Cardy:2016fqc, deBoer:2021zlm} \footnote{For a review, see \cite{Das:2024vqe}.}. The study of modular flow exhibits a generalization of thermal state to mixed state. In particular, as shown in \cite{Sorce:2023gio}, modular Hamiltonian is the unique self-adjoint operator which generates a thermal time evolution in out-of-equillibrium states\footnote{In the language of algebraic qft, these are cyclic and separating states with respect to an algebra $\mathcal{A}$, for which one could construct the modular operator.}. Hence in our context, studying quantization of modular Hamiltonian would shed light to address the key questions we mentioned.

In this paper, we closely follow the analysis by Tada \cite{Tada:2019rls} and develop it further to find the spectrum of this quantization and the corresponding bulk dual. The interesting feature of the quantization is the existence of fixed points along which the spatial coordinate ends. On the other hand, the (Euclidean) time forms circular trajectory which can be contracted to the fixed points. To get a well-defined and discrete Virasoro algebra, Tada introduced cut-off ($\epsilon$) around the fixed points. In this regulated Hilbert space, one can define Virasoro generators and their algebra with a distinct $\epsilon$ dependent central extension term. In this note, our main work is to construct the regulated Hilbert space $\mathcal{H}_{\epsilon}$ by finding the representation of the (modified) Virasoro algebra. The key step we use is to unwrap the (non-contractible) time circles in the presence of cut-off, which equivalently constructs a Lorentzian representation of the Euclidean quantization. This unwrapping will enable us to define a vacuum and highest weight primary states in the same spirit of state-operator correspondence in radial quantization. We also construct vacuum descendant state in this quantization. In $\epsilon \rightarrow 0$ limit, those (vacuum) descendants as well as all primaries will correspond to different vacuum solution of the original Hamiltonian. Taking the limit $\epsilon \rightarrow 0$ is determined by the choice of boundary condition at the cut-off. We will argue that we can write the vacuum of the regulated Hamiltonian as particular conformal boundary state. We also compute two point function in the $\epsilon \rightarrow 0$ limit in (vacuum) descendant and primary states. They will show thermal behavior in various limits. However, in contrast to radial quantization, two point function of light probes in heavy primary states of this quantization (with $h>c/24$) will show periodic behavior in Lorentzian time in the semiclassical limit. This is a manifestation of the fact that this modular quantization is unitarily inequivalent to radial quantization\cite{Das:2024vqe}. This inequivalence illustrates that the vacuum of the quantization in $\epsilon \rightarrow 0$ limit, which is eventually the same vacuum of radial quantization, must have a different bulk representation. In particular, we find that the bulk dual of the vacuum is described by AdS-Rindler or planar BTZ metric. We will further show the \textit{fixed points of the quantization correspond to the horizon(past and future) of the geometry}. This explains the cut-off around the fixed point will be  \textit{the natural bulk realization of stretched horizon.}  

Even though these observations give an answer to some of our main questions as we noted before, the emergence of type III algebra associated to smooth horizon is not yet addressed. This is because the regulated Hilbert space $\mathcal{H}_{\epsilon}$ is still an ad hoc construction by hand to make sense of a Virasoro algebra and to get a finite partition function. This is only physically meaningful if the states of $\mathcal{H}_{\epsilon}$ will contribute to the entropy of the quantization. Using the consistency condition of the regulated partition function, we obtain a thermal entropy at $\epsilon \rightarrow 0$ limit which agrees to the entanglement entropy of the boundary subregion or the thermal entropy associated to AdS-Rindler wedge. We also obtain high energy density of states and the corresponding microcanonical entropy of the quantization. Most notably, this microcanonical entropy gives the Bekenstein-Hawking entropy once we fix the energy to be the mass of the BTZ black hole. This also exactly reproduces the Hawking temperature. The non-trivial result of this computation is \textit{the high energy states which contribute to this entropy (or to the Cardy density of state) are the high energy states of the $\mathcal{H}_{\epsilon}$ at finite $\epsilon$. Thus the finite $\epsilon$ states of $\mathcal{H}_{\epsilon}$ are the microstates of BTZ in this description. In strict $\epsilon \rightarrow 0$ limit, the planar BTZ is described by the vacuum.} This would explain the meaning of \textit{emergent Hilbert space (or effective field theory fock space in the bulk) and the type III algebra of operators} at $\epsilon \rightarrow 0$ from a bigger Hilbert space $\mathcal{H}_{\epsilon}$ where the associated algebra will be of type I. We further conjecture the vacuum descendant states of $\mathcal{H}_{\epsilon}$ are dual to excited states (on top of the vacuum) of EFT Fock space built by free scalar fields in planar BTZ with stretched horizon background.

The rest of the paper is organized as follows: In \S\ref{sec2}, we will review the relevant part of Tada's construction \cite{Tada:2019rls} and construct the Hilbert space $\mathcal{H}_{\epsilon}$ and it's spectrum. In \S\ref{sec3} we compute two point functions in different eigenstates in different limits. We also argued the vacuum of $\mathcal{H}_{\epsilon}$ to be a conformal boundary state. Using the consistency condition of annulus partition function we compute both canonical and microcanonical entropy of the quantization in \S\ref{sec4}.
In \S\ref{sec5}, we discuss the dual AdS$_{3}$ description of different vacuum. The connection of fixed points to horizon and cut-off to stretched horizon will be established there. We also comment on the connection between stretched horizon story in planar BTZ to the vacuum descendants of the quantization. We end up with a detailed summary of our findings and some discussion of our results with some remarks on future direction in \S\ref{sec6}. In \S\ref{app1}, we provide some details of constructing a class of vacuum which gives a partial consistency check of the quantization we come up with.


\section{Modular quantization and spectrum}\label{sec2}

In Euclidean 2d CFT, the most famous and common notion of quantization is the so-called `radial quantization'. In this quantization scheme, CFT on a spatial cylinder ($S^{1}\times\mathbb{R}_{t}$) can be quantized by mapping it to the complex plane where the time flow ($t$) is generated radially. The most useful feature of this quantization is the manifestation of `state-operator correspondence' following which one can decompose the Hilbert space into irreducible Verma modules where all eigenstates can be obtained from all descendants of the highest weight primary \cite{Belavin:1984vu}. The Hamiltonian which generates such radial time flow is $L_{0}+\Bar{L}_{0}$ where $L_{n},\Bar{L}_{n}$ are modes of stress tensor. 

Here we consider a slightly different Hamiltonian generated by sl(2,$\mathbb{R}$) global conformal generators.
\begin{align}\label{Ham}
    H=\alpha (L_{0}+\Bar{L}_{0}) + \beta(L_{1}+L_{-1}+\Bar{L}_{1}+\Bar{L}_{-1}), \; \text{where} \; d \equiv \alpha^{2}-4\beta^{2} <0
\end{align}
The Hamiltonian is by construction Hermitian(i.e. $H=H^{\dagger}$) as in radial quantization $L_{n}^{\dagger}=L_{-n}$. This type of Hamiltonian appears to be the total modular Hamiltonian of a segment in a spatial circle \cite{Cardy:2016fqc, deBoer:2021zlm}. More generally, one can think of a CFT lived on a spatial circle $S^{1}$ for which the time flow is generated by (\ref{Ham}). When $d>0$, the Hamiltonian is unitarily equivalent to the Hamiltonian of radial quantization \cite{Das:2024vqe}. It has been studied in details in the context of `Mobius quantization' \cite{Okunishi:2016zat} where discrete Virasoro algebra as well as discrete spectrum is manifested. Another interesting class is when $d=0$ and $\beta=\frac{\alpha}{2}$ which is sometimes called `dipolar quantization' \cite{Ishibashi:2015jba}. This also has been studied for SSD Hamiltonian in great details in \cite{Ishibashi:2016bey} where a continuous Virasoro algebra emerges. The Hamiltonian with $d<0$ of the form (\ref{Ham}), has not been studied much unlike the other two classes\footnote{In the context of modular Hamiltonian in line interval, a Lorentzian Virasoro algebra has been obtained in \cite{Das:2020goe}. In fact, those Lorentzian Virasoro generators are isomorphic to the Euclidean modular Virasoro generators which we will introduce later in this section and this observation plays a crucial role to obtain the Lorentzian representation which we will describe here.}. In \cite{Tada:2019rls}, it was shown how a discrete Virasoro algebra could arise for these Hamiltonians after choosing a suitable cut-off \footnote{In \cite{deBoer:2021zlm}, it was argued that in the absence of any cut-off one could still get a continuous version of Virasoro algebra.}. In this note, we closely follow their work and set up. We dub quantization of (\ref{Ham}) as `modular quantization'. In this section, we begin by reviewing the technical part of \cite{Tada:2019rls} and proceed further to study the spectrum or highest-weight representation of the Virasoro algebra. To begin with, we briefly summarize the general rule of quantizing such class of sl(2,$\mathbb{R}$) deformed Hamiltonian following  \cite{Ishibashi:2016bey}.

\underline{\textit{Quantization contour:}}  We can rewrite (\ref{Ham}) as
\begin{align}
    H = \mathcal{L}_{0}+\Bar{\mathcal{L}}_{0}
\end{align}
The corresponding infinitesimal generators of $\mathcal{L}_{0}$ and $\mathcal{L}_{0}$ are $\Tilde{l}_{0}$ and $\Bar{\tilde{l}}_{0}$\footnote{For notational purpose, we denote the standard infinitesimal generators corresponding to Virasoro generators of radial quantization as $l_{n},\Bar{l}_{n}$.} where
\begin{align}
    \tilde{l}_{0}=g(z)\partial_{z}, \; \Bar{\tilde{l}}_{0}=\Bar{g}(\Bar{z})\partial_{\Bar{z}}
\end{align}
For (\ref{Ham}), $g(z)=\alpha z+\beta(z^{2}+1) \equiv \beta(z-z_{+})(z-z_{-})$. Here
\begin{align}
    z_{\pm} = \frac{-\alpha\pm i\sqrt{|d|}}{2\beta}
\end{align}
Hence $\Bar{z}_{+}=z_{-}, \Bar{z}_{-}=z_{+}$. From now on, we refer $d$ as $|d|$. The main idea \cite{Ishibashi:2016bey} of the sl(2,$\mathbb{R})$ quantization is that the conformal killing vector $(\tilde{l}_{0}+\bar{\tilde{l}}_{0})$ generates the new time direction $t$. Hence, the time flow is generated by $l_{0}+\Bar{l}_{0}$ and space direction is generated by $i(l_{0}-\Bar{l}_{0})$. One can show that \cite{Ishibashi:2016bey}
\begin{align}
    t+is = \int_{z}\frac{dz}{g(z)}, \; t-is = \int_{\Bar{z}}\frac{d\Bar{z}}{\Bar{g}(\Bar{z})}.
\end{align}
In our case,
\begin{align}\label{t,s curve}
    t+is = \frac{1}{i\sqrt{d}}\ln\frac{z-z_{+}}{z-z_{-}} \; \text{and} \; t-is = -\frac{1}{i\sqrt{d}}\ln\frac{\bar{z}-z_{-}}{\bar{z}-z_{+}}.
\end{align}
We will now find the constant $t$ and constant $s$ trajectory to determine the complex $z$ contour for this quantization. From (\ref{t,s curve}), we get for constant $t$
\begin{align}
    e^{2it\sqrt{d}} = \frac{(z-z_{+})(\bar{z}-z_{+})}{(z-z_{-})(\bar{z}-z_{-})}
\end{align}
If we denote $z=x+iy$ and $z_{\pm}=z^{R}_{*}\pm iz^{I}_{*}$, we will get
\begin{align}\label{A B}
    \tan \sqrt{d}t = \frac{B}{A}
\end{align}
where
\begin{align}
    &A=(x-z^{R}_{*})^{2}+(y^{2}-(z^{I}_{*})^{2}) \nonumber \\
    &B = -2z^{I}_{*}(x-z^{R}_{*})
\end{align}
Further simplifying (\ref{A B}), we have
\begin{align}
    (x-z^{R}_{*}+z^{I}_{*}\cot \sqrt{d}t)^{2}+y^{2} = (z^{I}_{*})^{2}(\csc\sqrt{d}t)^{2}
\end{align}
The constant $t$ curve seems to generate an equation of a circle. However one can notice that for $x=z_{\pm}$, the above equation satisfies trivially for all $t$. This implies the contour will pass through those points for all constant time $t$. In other words, $z_{\pm}$ are the fixed points of the contour(see Fig 2 of \cite{Tada:2019rls}).
Similarly, to find constant $s$ curve, we again use (\ref{t,s curve}) to get the following:
\begin{align}\label{s curve}
    e^{-2\sqrt{d}s} = \frac{(z-z_{+})(\bar{z}-z_{-})}{(z-z_{-})(\Bar{z}-z_{+})}
\end{align}
In a similar manner, the above equation can be simplified to get the following curve equation for time trajectory
\begin{align}
    (x-z^{R}_{*})^{2} + (y-z^{I}_{*}\coth\sqrt{d}s )^{2} = (z^{I}_{*})^{2}(\csch\sqrt{d}s)^{2}
\end{align}
Hence the constant $s$ curve is again a circle. Thus time flows in this quantization will generate circular trajectory. Interestingly, these are constractible time circle. In $s \rightarrow \pm \infty$ the circle reduces to a point. From (\ref{s curve})
 this corresponds to $z=z_{\pm}$. Hence, at the fixed point the time circle shrinks to zero size. Note that, in Euclidean AdS blackholes(Cigar geometry) there exists contractible time circles which smoothly shrink to it's origin where the horizon lies. In AdS/CFT, a well known fact is the existence of contractible time circle in the deconfined phase or black hole phase in large N phase transition \cite{Aharony:2003sx}.

\underline{\textit{Virasoro mode expansion:}} Following \cite{Ishibashi:2016bey}, we can define $f_{k}$ as the eigenmode of $\tilde{l}_{0}$
\begin{align}
    \tilde{l}_{0}f_{k}(z) = kf_{k}(z)
\end{align}
The solution for our case can be written as
\begin{align}\label{f}
    f_{k}(z)= e^{k\int \frac{dz}{g(z)}} = \left(\frac{z-z_{+}}{z-z_{-}}\right)^{-\frac{ik}{\sqrt{d}}}
\end{align}
Using this, one can construct the following modes
\begin{align}
    \tilde{l}_{k}(z)=g(z)f_{k}(z)\partial_{z}
\end{align}
It is easy to show that $\tilde{l}_{k}$ satisfy Witt algebra \cite{Ishibashi:2016bey}. Hence, one can define the conserved charges $\mathcal{L}_{k}$ corresponding to these conformal killing vectors $l_{k}$.
\begin{align}\label{vir gen}
    \mathcal{L}_{k} &= \frac{1}{2\pi i} \int_{C}dz g(z) f_{k}(z)T(z)\nonumber \\
    &= \frac{\beta}{2\pi i} \int_{C}dz (z-z_{+})^{-\frac{ik}{\sqrt{d}}+1}(z-z_{-})^{\frac{ik}{\sqrt{d}}+1} T(z)    
\end{align}
All the above discussion will also hold for anti-holomorphic modes $\Bar{\mathcal{L}}_{k}$. Here $C$ is the constant $t$ contour with all values of $s\in (-\infty,\infty)$. Using $TT$ ope it is straightforward to show that \cite{Tada:2019rls}
\begin{align}\label{Vir1}
    [\mathcal{L}_{k},\mathcal{L}_{k'}] = (k-k')\mathcal{L}_{k+k'} + \frac{c}{12} \frac{(k^{3}+k d)}{2\pi i}\int_{C}dz \frac{f_{k+k'}(z)}{g(z)}
\end{align}
By definition, one can change $z$ integration to $f_{k}$ integration
\begin{align}
    \frac{df_{k}}{f_{k}} = k\frac{dz}{g(z)}
\end{align}
Thus
\begin{align}\label{c integral}
   \int_{C}dz \frac{f_{k+k'}(z)}{g(z)} = \int_{C}\frac{df_{k+k'}}{(k+k')} = \frac{1}{(k+k')}(f_{k+k'}(z_{+})-f_{k+k'}(z_{-}))
\end{align}
However from (\ref{f}), we can easily see that $f_{k}(z_{\pm})$ are divergent and hence the central extension term of the algebra is not well-defined. This motivates \cite{Tada:2019rls} to place a cut-off around the fixed points $z=z_{\pm}$\footnote{Similar cut-off and boundary conditions at the entangling surfaces are previously discussed in \cite{Ohmori:2014eia, Cardy:2016fqc}.}
\begin{align}
    z=z_{\pm} + \epsilon e^{i\theta_{\epsilon}}
\end{align}
In the context of Rindler quantization of CFT or quantization in Minkowski space CFT, a notion called `angular quantization' is proposed in \cite{Agia:2022srj}, where a similar cut-off around fixed points is introduced to make the Hilbert space well-defined. In the presence of fixed points, one can not define a partition function (or traces). After introducing a cut-off and imposing suitable boundary condition on the cut-off one could define states in the (regulated) Hilbert space using the standard path integral on annulus. At the end, in the limit of removing the cut-off appropriately, the prescription offers to compute thermal correlators by the virtue of open-closed string duality in the annulus partition function. We will get back to this in the next section to connect modular quantization to angular quantization.  Around $z \sim z_{\pm}$, (\ref{t,s curve}) reduces to
\begin{align}
    (t+is)|_{z=z_{+}+\epsilon e^{i\theta_{\epsilon}}} &= \frac{1}{i\sqrt{d}}\ln\left(\frac{\epsilon e^{i\theta_{\epsilon}}}{i\sqrt{d}/\beta}\right) \nonumber \\
    &= \frac{\theta_{\epsilon}-\frac{\pi}{2}}{\sqrt{d}}+\frac{i}{\sqrt{d}}\ln\left(\frac{\sqrt{d}}{\beta\epsilon}\right)
\end{align}
Similarly
\begin{align}
    (t+is)|_{z=z_{-}+\epsilon e^{i\theta_{\epsilon}}} = \frac{-\theta_{\epsilon}-\frac{\pi}{2}}{\sqrt{d}}-\frac{i}{\sqrt{d}}\ln\left(\frac{\sqrt{d}}{\beta\epsilon}\right)
\end{align}
Hence in terms of $t,s$ one can rewrite $z=z_{\pm}+\epsilon e^{i\theta_{\epsilon}}$ as
\begin{align}\label{cutoff}
    z=z_{\pm} + e^{\mp \sqrt{d} s_{\epsilon} \pm i(\sqrt{d}t_{\epsilon}+\frac{\pi}{2})}
\end{align}
Here $s\rightarrow s_{\epsilon}$ and $t\rightarrow t_{\theta}$ on the cut-off. In terms of cut-off, we can have a finite integral (\ref{c integral}) and it will give
\begin{align}
    \int_{C}dz \frac{f_{k+k'}(z)}{g(z)} = \frac{e^{(k+k')t_{\epsilon}}}{(k+k')} \left[e^{i\frac{(k+k')}{\sqrt{d}}\ln\left(\frac{\sqrt{d}}{\beta\epsilon}\right)}-e^{-i\frac{(k+k')}{\sqrt{d}}\ln\left(\frac{\sqrt{d}}{\beta\epsilon}\right)}\right]
\end{align}
Putting this back into (\ref{Vir1}), the commutator\footnote{The modified generators with finite $\epsilon$ will be denoted as $\mathcal{L}^{\epsilon}_{k}$. The modification amounts a redefined constant $t$ contour where $s:\{ -\frac{1}{\sqrt{d}}\ln\left(\frac{\sqrt{d}}{\beta\epsilon}\right),\frac{1}{\sqrt{d}}\ln\left(\frac{\sqrt{d}}{\beta\epsilon}\right)\}$} becomes
\begin{align}\label{Vir2}
    [\mathcal{L}^{\epsilon}_{k},\mathcal{L}^{\epsilon}_{k'}] = (k-k')\mathcal{L}_{k+k'}^{\epsilon} + \frac{c}{12} \frac{(k^{3}+k d)e^{(k+k')t}}{\pi(k+k')}\sin\left(\frac{(k+k')}{\sqrt{d}}\ln\left(\frac{\sqrt{d}}{\beta\epsilon}\right)\right)
\end{align}
At this stage, the above algebra is true for any discrete or continuous label $k$. However, one could check that the above algebra does not close under Jacobi identity even though the Witt algebra (without the central extension term) satisfies it. In other word, the commutator is ill-defined due to the symmetric central extension part. To make a well-defined Virasoro algebra satisfying Jacobi identity, we must have the central extension term to be zero when $(k+k')\neq 0$. \footnote{If $\mathcal{L}_{m}$ satisfies the Virasoro algebra of the form $[\mathcal{L}_{m},\mathcal{L}_{n}]=(m-n)\mathcal{L}_{m+n}+C_{m,n}$, then the Jacobi identity for $\mathcal{L}_{m},\mathcal{L}_{n},\mathcal{L}_{p}$ implies:
$(m-n)C_{m+n,p}+(n-p)C_{n+p,m}+(p-m)C_{p+m,n}=0$. The central extension term of (\ref{Vir2}) does not satisfy this relation. To obtain such relation a Delta function will be needed.} This fixes a quantization of label $k$ in terms of integer $n$ as the following:
\begin{align}\label{quantization}
 k = \frac{n\pi \sqrt{d}}{\ln\left(\frac{\sqrt{d}}{\beta\epsilon}\right)}  , \; n\in \frac{\mathbb{Z}}{2}
\end{align}
It might be possible that $n \in \frac{\mathbb{Z}}{2}\; \text{or}\; \mathbb{Z}$. Remarkably, if $n \in \mathbb{Z}$, this has an exact similarity with normal modes of free scalar fields in the BTZ strecthed horizon background \cite{Burman:2023kko} as we will discuss in section \ref{sec5}. For some later purpose we just use $n \in \mathbb{Z}$. We can find the algebra for $(k+k') = 0$ by appropriately taking the limit in (\ref{Vir2})
\begin{align}\label{limit of delta}
    \lim_{(k+k')\rightarrow 0} \frac{(k^{3}+k d)e^{(k+k')t}}{\pi(k+k')}\sin\left(\frac{(k+k')}{\sqrt{d}}\ln\left(\frac{\sqrt{d}}{\beta\epsilon}\right)\right) = \frac{(k^{3}+k d)}{\pi\sqrt{d}}\ln\left(\frac{\sqrt{d}}{\beta\epsilon}\right)
\end{align}
Hence the final form of the modified Virasoro algebra is
\begin{align}\label{Vir algebra}
    [\mathcal{L}^{\epsilon}_{k},\mathcal{L}_{k'}^{\epsilon}] &= (k-k')\mathcal{L}_{k+k'}^{\epsilon} + \frac{c}{12}\frac{(k^{3}+k d)}{\pi\sqrt{d}}\ln\left(\frac{\sqrt{d}}{\beta\epsilon}\right) \delta_{k+k'} \\
  & =   (k-k')\mathcal{L}_{k+k'}^{\epsilon} + \frac{c_{eff}}{12}(k^{3}+k d) \delta_{k+k'}
\end{align}
Where we define $c_{eff} = \frac{c}{\pi\sqrt{d}}\ln\left(\frac{\sqrt{d}}{\beta\epsilon}\right)$. In $\epsilon \rightarrow 0$ limit (with finite $c$), $c_{eff}\rightarrow \infty$. Interestingly, in the same limit $k$ becomes continuous from (\ref{quantization}) and a continuous dirac delta function emerges\cite{deBoer:2021zlm}\footnote{To see this one should first take the $\epsilon \rightarrow 0$ limit in (\ref{limit of delta}) before taking $k+k' \rightarrow 0$.}. Hence $\epsilon \rightarrow 0$ limit gives a continuous non-normalizable spectrum with continuous Virasoro algebra. In section \ref{sec5} we will see a direct correspondence with this fact to strecthed horizon quantization in AdS$_{3}$ where continuous spectrum emerges at the limit of approaching the stretched horizon to horizon. However we should note that the central extension term is not exactly similar to usual Virasoro central term. The difference lies in $k^{2}(k+d)$ factor. It reduces to usual Virasoro only if $d=-1$ which is not possible in our case. This has an important consequence in computing descendant correlators which we will discuss in the next section.

\underline{\textit{Hermitian conjugate:}} In this new quantization an important task is to understand the Hermitian conjugate of the field operators, stress tensors as well as the Virasoro generators. We follow \cite{Ishibashi:2016bey} for constructing Hermitian conjugate in this new quantization. In Lorentzian time $\tau$, a real operator in certain spacetime point is manifestly Hermitian since $\phi^{\dagger}(0)=\phi(0)$ and as we mentioned earlier $H=H^{\dagger}$:
\begin{align}
    \phi(\tau)^{\dagger} = (e^{iH\tau}\phi(0)e^{-iH\tau})^{\dagger} = \phi(\tau)
\end{align}
However in Euclidean case $\phi^{\dagger}(t)=(e^{Ht}\phi(o)e^{-Ht})^{\dagger} = \phi(-t)$. Thus in Euclidean case, taking Hermitian conjugate flips $t \rightarrow -t$. This further implies $\omega(=t+is) \rightarrow -\bar{\omega}$. If $\phi(z)$ is a primary scalar\footnote{For simplicity we consider only holomorphic sector. The whole analysis can be easily generalized to generic primary fields having both holomorphic and anti-holomorphic part.} under radial quantization with dimension $h$, then we define $\phi'(\omega) = \left(\frac{\partial\omega}{\partial z}\right)^{-h}\phi(z)$. Hence Hermitian conjugate in our sl(2,$\mathbb{R}$) quantization implies:
\begin{align}\label{herm1}
    (\phi'(\omega))^{\dagger} = \phi'(-\bar{\omega})
\end{align}
From (\ref{t,s curve}), $\omega = t+is =\frac{1}{i\sqrt{d}}\ln\frac{z-z_{+}}{z-z_{-}}$. We can invert this to get
\begin{align}\label{z s transf}
    &z-z_{-} = \frac{i\sqrt{d}}{\beta}\frac{1}{1-e^{i\omega\sqrt{d}}} \\
    & z-z_{+} = \frac{i\sqrt{d}}{\beta}\frac{e^{i\omega\sqrt{d}}}{1-e^{i\omega\sqrt{d}}}
\end{align}
Similarly
\begin{align}
    &\bar{z}-z_{+} = \frac{i\sqrt{d}}{\beta}\frac{e^{i\Bar{\omega}\sqrt{d}}}{1-e^{i\bar{\omega}\sqrt{d}}} \\
    &\bar{z}-z_{-} = \frac{i\sqrt{d}}{\beta}\frac{1}{1-e^{i\bar{\omega}\sqrt{d}}}
\end{align}
From the primary field transformation, we have
\begin{align}
    \phi'(\omega) =\left(\frac{1}{\beta(z-z_{+})(z-z_{-})}\right)^{-h} \phi(z) = \left[-\frac{\beta(1-e^{i\omega\sqrt{d}})^{2}}{de^{i\omega\sqrt{d}}}\right]^{-h}\phi\left(\frac{i\sqrt{d}}{\beta}\frac{1}{1-e^{i\omega\sqrt{d}}}+z_{-}\right)
\end{align}
From the above expression we get the Hermitian conjugate of $\phi'(\omega)$ as
\begin{align}\label{herm2}
    (\phi'(\omega))^{\dagger} = \left[-\frac{\beta(1-e^{-i\bar{\omega}\sqrt{d}})^{2}}{de^{-i\bar{\omega}\sqrt{d}}}\right]^{-h} (\phi(z))^{\dagger}
\end{align}
Also we get
\begin{align}\label{herm3}
    \phi'(-\bar{\omega}) &= \left[-\frac{\beta(1-e^{-i\bar{\omega}\sqrt{d}})^{2}}{de^{-i\bar{\omega}\sqrt{d}}}\right]^{-h} \phi\left(\frac{i\sqrt{d}}{\beta}\frac{1}{1-e^{-i\bar{\omega}\sqrt{d}}}+z_{-}\right) \\
    & = \left[-\frac{\beta(1-e^{-i\bar{\omega}\sqrt{d}})^{2}}{de^{-i\bar{\omega}\sqrt{d}}}\right]^{-h} \phi(z_{+}+z_{-}-\bar{z})
\end{align}
Hence from (\ref{herm1}), (\ref{herm2}) and (\ref{herm3}) we finally get
\begin{align}
    \phi^{\dagger}(z) = \phi(z_{+}+z_{-}-\bar{z})
\end{align}
This can be checked to be true for stress tensor i.e.
\begin{align}\label{herm4}
    T^{\dagger}(z) = T(z_{+}+z_{-}-\bar{z})
\end{align}
We will now check the Hermiticity condition on conformal generators $\mathcal{L}_{k}$ \footnote{The same discussion is also applicable to discrete Virasoro generators $\mathcal{L}^{\epsilon}_{k}$}. From the definition
\begin{align}
    \mathcal{L}_{k} = \frac{\beta}{2\pi i} \int_{C} dz (z-z_{+})^{-\frac{ik}{\sqrt{d}}+1}(z-z_{-})^{\frac{ik}{\sqrt{d}}+1} T(z)
\end{align}
Hence using (\ref{herm4}) we get
\begin{align}
    \mathcal{L}^{\dagger}_{k} = -\frac{\beta}{2\pi i} \int_{C} d\bar{z} (\bar{z}-z_{-})^{\frac{ik}{\sqrt{d}}+1}(\bar{z}-z_{+})^{-\frac{ik}{\sqrt{d}}+1} T(z_{+}+z_{-}-\bar{z})
\end{align}
Using a change of variable $\bar{z'} = (z_{+}+z_{-}-\bar{z})$, we get the following
\begin{align}
     \mathcal{L}^{\dagger}_{k} = -\frac{\beta}{2\pi i} \int_{C} d\bar{z'}(\bar{z'}-z_{+})^{\frac{ik}{\sqrt{d}}+1}(\bar{z'}-z_{-})^{-\frac{ik}{\sqrt{d}}+1} T(\bar{z'})
\end{align}
In the above, one negative sign comes from the substitution of new variable and another negative sign appears due to change of direction of contour. Hence the overall sign remains the same. Now if we change the above integral from anti-holomorphic ($\bar{z'}$) to holomorphic $(\hat{z})$ another negative sign will appear due to change in the direction of the contour. So, finally we arrive at:
\begin{align}
     \mathcal{L}^{\dagger}_{k} &= \frac{\beta}{2\pi i} \int_{C} d\hat{z}(\hat{z}-z_{+})^{\frac{ik}{\sqrt{d}}+1}(\hat{z}-z_{-})^{-\frac{ik}{\sqrt{d}}+1} T(\hat{z}) \nonumber \\
     &=\mathcal{L}_{-k}
\end{align}
This is similar to the Hermiticity properties of $L_{n}$ in radial quantization. 

\underline{\textit{Spectrum:}} The appearance of discrete Virasoro algebra made out of $\mathcal{L}^{\epsilon}_{k}$ is encouraging to define it's highest weight representation. This would be similar to the construction of Verma module in radial quantization. However, the only obstacle in the sl(2,$\mathbb{R}$) quantization we consider is that the time direction is compact unlike in radial quantization. Hence in our setting, thermal trace can be constructed naturally \cite{Agia:2022srj} in the path integral picture where periodic boundary condition in the temporal direction of fields has been imposed. In the next section, we will discuss some aspects of it to compute correlation function in this quantization. However, one may still wonder whether we could construct a different meaningful irreducible representation of the Virasoro algebra we have after placing a cut-off near the fixed points. Without a cut-off, the time circles are contractible as we mentioned earlier. Since those circles can be shrunk to the fixed points continuously, those are simply connected curves.  However after placing a cut-off the circles are not contractible to the fixed point. Hence in the cut-off case we can always find an universal cover which is a real line. In other words, we can unwrap the non-contractible time circles to a line which is meaningful to construct a ground state of the quantization. This implies we take $t:(-\infty,\infty)$ in this regularized setting \footnote{One instance for such unwrapping is discussed in the context of getting AdS hyperboloid by embedding it in the higher dimension \cite{Aharony:1999ti}. In this embedding mechanicsm, one could get a global AdS with periodic time direction in Lorentzian. To avoid closed time-like curves, one should unwrap it and find an universal covering of AdS.}. We would like to view this stretching of time or going to the universal cover as the intermediate state from Euclidean to Lorentzian\footnote{The author would like to thank Bobby Ezhuthachan and Chethan Krishnan for emphasizing the importance to view this as Lorentzian}. We observe that upon analytic continuation $t \rightarrow it_{L}$ and by changing the label of Virasoro generator $k \rightarrow ik$ and $d \rightarrow -d$ one could obtain a Lorentzian modes where the integration range of $z,\Bar{z} \in (-\infty,\infty)$\footnote{Here $(z,\Bar{z}) \equiv (t+s,t-s)$.} which will satisfy the standard Virasoro algebra as in \cite{Das:2020goe}. It is easy to check that all of our subsequent analysis of constructing spectrum will also hold for this Lorentzian setting. This makes an equivalence between Euclidean and Lorentzian CFTs in this quantization. In other way, we should think all of the succeeding analysis as the Lorentzian construction of highest weight representation in Euclidean thermal or modular quantization. We will first define vacuum, in and out state in the same spirit of radial quantization by taking $t \rightarrow -\infty$ limit \footnote{Though the spirit constructing the Hilbert space is similar to angular quantization as described in \cite{Agia:2022srj}, our approach is different from them where we exploit Virasoro algebra and it's above-mentioned Lorentzian representation to define a Hilbert space.}. Note that, to make it consistent with the Lorentzian analysis, we stretched $t$ direction from one cut-off$(z_{+)}$ to another one $z_{-}$. Thus we have $t_{\epsilon} = -\infty$.

Firstly, we want to construct a mode expansion of $T(z)$ in terms of $\mathcal{L}^{\epsilon}_{k}$s as the following:
\begin{align}\label{T ansatz}
    T(z) = \frac{\pi \sqrt{d}}{\beta^{2}\ln\left(\frac{\sqrt{d}}{\beta\epsilon}\right)}\sum_{k} (z-z_{+})^{\frac{ik}{\sqrt{d}}-2}(z-z_{-})^{-\frac{ik}{\sqrt{d}}-2}\mathcal{L}^{\epsilon}_{k}
\end{align}
To check this, we use (\ref{vir gen}) in (\ref{T ansatz}) to obtain the following:
\begin{align}
    \mathcal{L}^{\epsilon}_{k} = \frac{\beta}{2\pi i}\frac{\pi \sqrt{d}}{\beta^{2}\ln\left(\frac{\sqrt{d}}{\beta\epsilon}\right)}\sum_{k'}\int_{C'}dz \left(\frac{z-z_{+}}{z-z_{-}}\right)^{\frac{i(k'-k)}{\sqrt{d}}} \frac{1}{(z-z_{+})(z-z_{-})}\mathcal{L}^{\epsilon}_{k'}
\end{align}
We may choose constant $t=0$ contour for simplification and the integral becomes

$\int_{C'}dz \rightarrow -\frac{id}{\beta}\int^{\frac{1}{\sqrt{d}}\ln\left(\frac{\sqrt{d}}{\beta\epsilon}\right)}_{-\frac{1}{\sqrt{d}}\ln\left(\frac{\sqrt{d}}{\beta\epsilon}\right)}ds \frac{e^{\sqrt{d}s}}{(e^{\sqrt{d}s}-1)^{2}}$. Thus by using (\ref{z s transf}) we finally have
\begin{align}
   \mathcal{L}^{\epsilon}_{k} = \frac{\sqrt{d}}{2i\ln\left(\frac{\sqrt{d}}{\beta\epsilon}\right)} \sum_{k'} \int^{\frac{1}{\sqrt{d}}\ln\left(\frac{\sqrt{d}}{\beta\epsilon}\right)}_{-\frac{1}{\sqrt{d}}\ln\left(\frac{\sqrt{d}}{\beta\epsilon}\right)}ds e^{i(k'-k)s} \mathcal{L}^{\epsilon}_{k'}
\end{align}
Using (\ref{quantization}) one can immediately see that the integral becomes a delta function $\delta_{k-k'}$ when $n \in \mathbb{Z}$\footnote{For $n \in \frac{\mathbb{Z}}{2}$, in the definition of (\ref{T ansatz}) there will be a change of $k \rightarrow 2k$. This does not affect the rest of the analysis.} and hence the
sum picks up $k=k'$. To define a vacuum $|0_{\epsilon}\rangle$
 we use the fact that $T(z)|0_{\epsilon}\rangle$ to be well-defined at $t \rightarrow -\infty$. From (\ref{T ansatz})
 \begin{align}
     T(z) = \frac{\pi\beta^{2}}{d\sqrt{d}\ln\left(\frac{\sqrt{d}}{\beta\epsilon}\right)}\sum_{k} e^{-k(t+is)} \frac{\beta^{2}(1-e^{i\sqrt{d}(t+is)})^{4}}{de^{2i(t+is)\sqrt{d}}}\mathcal{L}^{\epsilon}_{k}
 \end{align}
If we now take $t\rightarrow -\infty$ limit, we have:
 \begin{align}
     \lim_{t \rightarrow -\infty}T(z)|0_{\epsilon}\rangle = \lim_{t\rightarrow -\infty}\frac{\pi\beta^{2}}{d\sqrt{d}\ln\left(\frac{\sqrt{d}}{\beta\epsilon}\right)}\sum_{k} e^{-k(t+is)} \frac{\beta^{2}(1-e^{i\sqrt{d}(t+is)})^{4}}{de^{2i(t+is)\sqrt{d}}} \mathcal{L}^{\epsilon}_{k} |0_{\epsilon}\rangle
 \end{align}
 The above limit suggests that to make a well-defined vacuum we must have the following
 \begin{align}\label{vac cond1}
     \mathcal{L}^{\epsilon}_{k}|0_{\epsilon}\rangle = 0 \; \text{for} \; k \geq 0
 \end{align}
However it is not clear whether this vacuum coincides with the vacuum defined in radial quantization. To check this we consider the mode expansion of $T(z)$ in radial quantization around $z=z_{-}$ in the definition of $\mathcal{L}_{k}^{\epsilon}$.
\begin{align}
    \mathcal{L}^{\epsilon}_{k} = \frac{\beta}{2\pi i}\int_{C'}dz (z-z_{+})^{-\frac{ik}{\sqrt{d}}+1}(z-z_{-})^{\frac{ik}{\sqrt{d}}+1}\sum_{m}(z-z_{-})^{-m-2}L_{m}
\end{align}
 Again using (\ref{z s transf}) and taking $t=0$ the above integral reduces to
 \begin{align}\label{bekar}
      \mathcal{L}^{\epsilon}_{k} = -\left(\frac{i\sqrt{d}}{\beta}\right)^{-m+1}\frac{\sqrt{d}}{2\pi}\sum_{m}\int^{\frac{1}{\sqrt{d}}\ln\left(\frac{\sqrt{d}}{\beta\epsilon}\right)}_{-\frac{1}{\sqrt{d}}\ln\left(\frac{\sqrt{d}}{\beta\epsilon}\right)}ds \frac{e^{-2\sqrt{d}s+iks}}{(1-e^{-\sqrt{d}s})^{-m+2}}L_{m}
 \end{align}
 This integral is hard to compute. However we can take $\epsilon \rightarrow 0$ limit and the limit of $s$ becomes $(-\infty,\infty)$. Using (\ref{vac cond1}), (\ref{bekar}) reduces to at $\epsilon \rightarrow 0$:
 \begin{align}
   \lim_{\epsilon\rightarrow 0}\mathcal{L}^{\epsilon}_{k>0}|0_{\epsilon}\rangle =  -\left(\frac{i\sqrt{d}}{\beta}\right)^{-m+1}\frac{\sqrt{d}}{2\pi}\sum_{m>1} \frac{2\Gamma(m-1)\Gamma\left(2-\frac{ik}{\sqrt{d}}\right)}{\sqrt{d}\Gamma\left(1+m-\frac{ik}{\sqrt{d}}\right)}L_{m}|0_{\epsilon\rightarrow 0}\rangle = 0
 \end{align}
 This implies $L_{m}|0_{\epsilon\rightarrow 0}\rangle = 0$ for $m>1$. If we expand $T(z)$ in radial quantization around $z=z_{+}$ in the definition of $\mathcal{L}_{k}^{\epsilon}$ we will get
 \begin{align}
     \mathcal{L}^{\epsilon}_{k} = \frac{\beta}{2\pi i}\int_{C'}dz (z-z_{+})^{-\frac{ik}{\sqrt{d}}+1}(z-z_{-})^{\frac{ik}{\sqrt{d}}+1}\sum_{m}(z-z_{+})^{-m-2}L_{m}
\end{align}
 In a similar way using (\ref{z s transf}) and taking $t=0$ and $\epsilon \rightarrow 0$ limit, the above integral yields
 \begin{align}\label{vac cond2}
      \mathcal{L}^{\epsilon}_{k} &= -\left(\frac{i\sqrt{d}}{\beta}\right)^{-m+1}\frac{\sqrt{d}}{2\pi}\sum_{m}\int^{\infty}_{-\infty}ds \frac{e^{m\sqrt{d}s+iks}}{(1-e^{-\sqrt{d}s})^{-m+2}}L_{m} \nonumber \\
      &=-\left(\frac{i\sqrt{d}}{\beta}\right)^{-m+1}\frac{\sqrt{d}}{2\pi}\sum_{m>1} \frac{2\Gamma(m-1)\Gamma\left(-m-\frac{ik}{\sqrt{d}}\right)}{\sqrt{d}\Gamma\left(-1-\frac{ik}{\sqrt{d}}\right)}L_{m}|0_{\epsilon\rightarrow 0}\rangle = 0
 \end{align}
Hence, again we have  $L_{m}|0_{\epsilon\rightarrow 0}\rangle = 0$ for $m>1$. This is true even for vacuum $|0\rangle$ as well as any other primary states $|h\rangle$ constructed in radial quantization. Thus we have seen $|0_{\epsilon \rightarrow 0}\rangle = |0\rangle$ or $|h\rangle$. This degeneracy of the choice of vacuum in $\epsilon \rightarrow 0$ suggests we do not expect $|0_{\epsilon \neq 0}\rangle$ is same to be $|0\rangle$. The choice of vacuum in $\epsilon \rightarrow 0$ is determined from the boundary condition we choose to prepare the state $|0_{\epsilon}\rangle$ in path integral. In the context of angular quantization \cite{Agia:2022srj}, this shrinking of boundary condition was studied in details to construct thermal correlators in that quantization. In that case, it was shown that the shrinking of boundary condition is not unique and the boundary condition will shrink to an operator at $\epsilon \rightarrow 0$ with lowest operator dimension (which may include idenitity operator) in radial quantization. In our construction, we will argue that the choice of boundary condition that leads to $|0_{\epsilon \rightarrow 0}\rangle \rightarrow |0\rangle$ is completely meaningful and well-motivated. This fact will be crucial while computing correlators as in the next section in $\epsilon \rightarrow 0$ limit. We will get back to the construction of $|0_{\epsilon}\rangle$ in the next section while connecting this quantization to angular quantization. We will argue that $|0_{\epsilon}\rangle$ to be a conformal boundary state. Our next goal is to find the primary states in this regulated representation of Virasoro algebra. To do this, we first compute the action of $\mathcal{L}^{\epsilon}_{k}$ on primary field (in radial quantization) $\phi_{h,\Bar{h}}(z,\Bar{z})$. Using (\ref{vir gen}) and $T\phi$ ope we have
 \begin{align}
     &[\mathcal{L}^{\epsilon}_{k},\phi(z,\Bar{z})] \nonumber \\
 &=\frac{\beta}{2\pi i}\oint_{z}d\omega (\omega-z_{+})^{-\frac{ik}{\sqrt{d}}+1}(\omega-z_{-})^{\frac{ik}{\sqrt{d}}+1}T(\omega)\phi(z,\bar{z}) \nonumber \\
 &=\frac{\beta}{2\pi i}\oint_{z}d\omega (\omega-z_{+})^{-\frac{ik}{\sqrt{d}}+1}(\omega-z_{-})^{\frac{ik}{\sqrt{d}}+1} \left[\frac{h\phi(z,\bar{z})}{(\omega-z)^{2}}+\frac{\partial_{z}\phi(z,\bar{z})}{\omega-z}\right] \nonumber \\
 &=(2\beta z+ \frac{\alpha+k}{\beta})(z-z_{+})^{-\frac{ik}{\sqrt{d}}}(z-z_{-})^{\frac{ik}{\sqrt{d}}}h\phi(z,\Bar{z}) + \beta (z-z_{+})^{-\frac{ik}{\sqrt{d}}+1}(z-z_{-})^{\frac{ik}{\sqrt{d}}+1} \partial_{z}\phi(z,\Bar{z})
 \end{align}
 We now want to evaluate the commutator at cut-off point $z=z_{\pm}+\epsilon e^{i\theta_{\epsilon}}$. Since $\theta_{\epsilon}$ explicitly related to $t_{\epsilon} =-\infty$, the states will be only meaningful near the cut-off. 
 At this point
 \begin{align}\label{primary1}
[\mathcal{L}^{\epsilon}_{k},\phi(z,\Bar{z})]|_{z=z_{+}+\epsilon e^{i\theta_{\epsilon}}} &= (2i\sqrt{d}+n+2\epsilon e^{i\theta_{\epsilon}})(\epsilon e^{i\theta_{\epsilon}})^{-\frac{ik}{\sqrt{d}}}\left(\frac{i\sqrt{d}}{\beta}+\epsilon e^{i\theta_{\epsilon}}\right)^{\frac{ik}{\sqrt{d}}}h\phi(z,\bar{z})|_{z=z_{+}+\epsilon e^{i\theta_{\epsilon}}} + \nonumber \\
&+\beta (\epsilon e^{i\theta_{\epsilon}})^{-\frac{ik}{\sqrt{d}}+1}\left(\frac{i\sqrt{d}}{\beta}+\epsilon e^{i\theta_{\epsilon}}\right)^{\frac{ik}{\sqrt{d}}+1} \partial_{z}\phi(z,\Bar{z})|_{z=z_{+}+\epsilon e^{i\theta_{\epsilon}}} 
 \end{align}
Using (\ref{cutoff}) , we rewrite $z_{+}+\epsilon e^{i\theta_{\epsilon}} = z_{+}+ e^{ -\sqrt{d} s_{\epsilon} + i(\sqrt{d}t_{\epsilon}+\frac{\pi}{2})}$. If we compute the action of the commutator on $|0_{\epsilon}\rangle$ then we must take $t \rightarrow -\infty = t_{\epsilon}$. Thus we get
\begin{align}
&[\mathcal{L}^{\epsilon}_{k},\phi(z,\Bar{z})]|_{z=z_{+}+\epsilon e^{i\theta_{\epsilon}}} |0_{\epsilon}\rangle = \lim_{t \rightarrow -\infty}  [\mathcal{L}^{\epsilon}_{k},\phi(z,\Bar{z})]|_{z=z_{+}+ e^{ -\sqrt{d} s + i(\sqrt{d}t+\frac{\pi}{2})}}|0_{\epsilon}\rangle \nonumber \\
  & = \lim_{t\rightarrow -\infty} [(2i\sqrt{d}+n+2\epsilon e^{i\theta_{\epsilon}})(e^{ -\sqrt{d} s + i(\sqrt{d}t+\frac{\pi}{2})})^{-\frac{ik}{\sqrt{d}}}\left(\frac{i\sqrt{d}}{\beta}+e^{ -\sqrt{d} s + i(\sqrt{d}t+\frac{\pi}{2})}\right)^{\frac{ik}{\sqrt{d}}}h\phi(z,\bar{z})|_{z=z_{+}+\epsilon e^{i\theta_{\epsilon}}} + \nonumber \\
&+\beta (e^{ -\sqrt{d} s + i(\sqrt{d}t+\frac{\pi}{2})})^{-\frac{ik}{\sqrt{d}}+1}\left(\frac{i\sqrt{d}}{\beta}+e^{ -\sqrt{d} s + i(\sqrt{d}t+\frac{\pi}{2})}\right)^{\frac{ik}{\sqrt{d}}+1} \partial_{z}\phi(z,\Bar{z})|_{z=z_{+}+\epsilon e^{i\theta_{\epsilon}}} ]|0_{\epsilon}\rangle
\end{align}
One can easily check that for $k>0$, the above expression vanishes. However for $k<0$, the above expression diverges. Hence the descendants of the primaries do not exist. The same conclusion holds also for another fixed point $z=z_{-}+\epsilon e^{i\theta_{\epsilon}} = z_{-}+ e^{ \sqrt{d} s_{\epsilon} - i(\sqrt{d}t_{\epsilon}+\frac{\pi}{2})}$ Also for $k>0$, $\mathcal{L}^{\epsilon}_{k}$ annihilates the vacuum. Hence, this constructs an in state in our quantization.
\begin{align}\label{in}
    |\phi_{in}^{\epsilon}\rangle \equiv |h,\bar{h}\rangle^{\epsilon} = \lim_{t \rightarrow -\infty}\phi_{h,\bar{h}}(z,\bar{z}) |0_{\epsilon}\rangle|_{z=z_{\pm}+\epsilon e^{i\theta_{\epsilon}}}
\end{align}
In a similar way, we could also construct $\langle \phi_{out}|$. Since $\phi^{\dagger}(z)=\phi(z_{+}+z_{-}-z)$ and $(\mathcal{L}_{n}^{\epsilon})^{\dagger}=\mathcal{L}_{-n}^{\epsilon}$, we can show that
\begin{align}\label{out}
    \langle \phi_{out}| \equiv {}^{\epsilon}\langle  h,\bar{h}| = (|h,\bar{h}\rangle^{\epsilon})^{\dagger}
\end{align}
To complete the highest-weight representation we must show $\mathcal{L}^{\epsilon}_{0}|h\rangle = h|h\rangle$. (\ref{primary1}) yields 
\begin{align}
&\lim_{t \rightarrow -\infty}[\mathcal{L}^{\epsilon}_{0},\phi(z,\Bar{z})]|_{z=z_{\pm}+\epsilon e^{i\theta_{\epsilon}}}  \nonumber \\
& = \lim_{t \rightarrow -\infty} (i\sqrt{d} + 2\beta\epsilon e^{i\theta_{\epsilon}}) h\phi(z,\bar{z})|_{z=z_{\pm}+\epsilon e^{i\theta_{\epsilon}}}
\end{align}
The complex multiplication factor makes it non-unitary\footnote{This is reminiscent of the fact that due to conformal boundary condition on cut-off there should be only one copy of Virasoro algebra \cite{Das:2024vqe}. Even though we primarily constructed holomorphic and anti holomorphic Viraoso generators, those two are not independent.} However if we consider total Hamiltonian $\mathcal{L}^{\epsilon}_{0}+\bar{\mathcal{L}}^{\epsilon}_{0}$ and consider scalar primary operator with $h=\Bar{h}$, we have
\begin{align}
   &\lim_{t \rightarrow -\infty}[\mathcal{L}^{\epsilon}_{0}+\bar{\mathcal{L}}^{\epsilon}_{0},\phi(z,\Bar{z})]|_{z=z_{\pm}+\epsilon e^{i\theta_{\epsilon}}}  \nonumber \\ 
   &=\lim_{t\rightarrow -\infty}4\beta\epsilon\cos\theta_{\epsilon} h\phi(z,\bar{z})|_{z=z_{\pm}+\epsilon e^{i\theta_{\epsilon}}}
\end{align}
This along with the fact $\mathcal{L}^{\epsilon}_{0}|0_{\epsilon}\rangle = 0$  suggests
\begin{align}\label{new vac}
    (\mathcal{L}^{\epsilon}_{0}+\bar{\mathcal{L}}^{\epsilon}_{0})|h,h\rangle^{\epsilon} = 4\beta\epsilon\cos\theta_{\epsilon} h|h,h\rangle^{\epsilon}
\end{align}
Note that, in $\epsilon \rightarrow 0$ the above expression vanishes for any finite $h$. In that limit, $|h,h\rangle$ itself behaves like another vacuum under this quantization. This fact can be checked independently as shown in appendix \ref{app1}. In other words, to be consistent in this quantization there should not exist operators with dimension $h$ such that $h\epsilon \neq 0$ when $\epsilon \rightarrow 0$. Even though, there do not exist any well-defined Virasoro descendants of primaries $|h,h\rangle^{\epsilon}$ in this quantization, one still construct vacuum descandents of the form $\prod_{\{n_{i},k_{i}\}}(\mathcal{L}^{\epsilon}_{-k_{1}})^{n_{1}}(\mathcal{L}^{\epsilon}_{-k_{2}})^{n_{2}}\dots (\bar{\mathcal{L}}^{\epsilon}_{-k_{1}})^{n_{1}}(\bar{\mathcal{L}}^{\epsilon}_{-k_{2}})^{n_{2}}\dots|0_{\epsilon}\rangle$. Hence all the eigenstates of $H_{\epsilon}$ we constructed so far, have non-vanishing energy at finite $\epsilon$. This fact has a parallel to stretched horizon quantization story where excited states on top of stretched horizon vacuum may have an one-to-one correspondence to the (vacuum) descendant states in this quantization. In the next section we will compute two point function in different eigenstates of this quantization.

\section{Eigenstate two point function}\label{sec3}

So far we have identified the eigenstates of the modular Hamiltonian in the regulated Hilbert space $\mathcal{H}_{\epsilon}$. This consists of vacuum $|0_{\epsilon}\rangle$, descendants of vacuum $\prod_{n_{i},k_{i}}(\mathcal{L}^{\epsilon}_{-k_{i}})^{n_{i}}|0_{\epsilon}\rangle$ and scalar primaries $|h,h\rangle^{\epsilon}$. In this section, we will compute two point functions in these eigenstates in certain limiting cases.

\underline{\textit{Descendant two point function:}} An interesting distinction between regulated modular quantization and radial quantization is the role played by $L_{\pm 1}$ generators. In radial quantization the commutator $[L_{1},L_{-1}]=2L_{0}$ remains $c$-independent and hence contributes to global Virasoro block at large $c$. However from (\ref{Vir algebra}) we can see that there exists no $k$ for which the central extension term will vanish.
Thus the norm of any descendant states  will be $c_{eff}$ dependent. Now we consider two point function in a generic descendant state of level $N_{lev}$. The states we consider are of the form
\begin{align}\label{descendant level}
  |\psi\rangle^{\epsilon} \equiv  \prod_{k,k'}(\mathcal{L}^{\epsilon}_{-k})^{N_{k}}(\bar{\mathcal{L}}^{\epsilon}_{-k'})^{N_{k'}}|0_{\epsilon}\rangle, \; \text{with} \; \sum_{k}kN_{k} + \sum_{k'}k' N_{k'} = N_{lev}.
\end{align}
The total conformal dimension of the state is $N_{lev}$. Without loss of generality we consider the following normalized two point function of probe light operator $O_{L}$ of dimension at $(z,\Bar{z}$ and $z=0,\Bar{z}=0$\footnote{$(z,\Bar{z})=(0,0)$ correspond to $t=0,s=0$ when $\alpha=0$. This is always possible since even for $\alpha=0$ and $\beta \neq 0$ the Hamiltonian belongs to the same class with $d<0$.} in the heavy descendant state $|\psi\rangle^{\epsilon}$ which we constructed above:
\begin{align}\label{desc corr}
   & \frac{{}^{\epsilon}\langle\psi|O_{L}(z,\bar{z})O_{L}(1,1)|\psi\rangle^{\epsilon}}{{}^{\epsilon}\langle\psi|\psi\rangle^{\epsilon}} \nonumber \\
   &= \frac{\langle 0_{\epsilon}|\left[\prod_{k,k'}(\mathcal{L}^{\epsilon}_{k})^{N_{k}}(\bar{\mathcal{L}}^{\epsilon}_{k'})^{N_{k'}},O_{L}(z,\Bar{z})\right]O_{L}(0,0)\prod_{k,k'}(\mathcal{L}^{\epsilon}_{-k})^{N_{k}}(\bar{\mathcal{L}}^{\epsilon}_{-k'})^{N_{k'}}|0_{\epsilon}\rangle}{\langle 0_{\epsilon}|\left[\prod_{k,k'}(\mathcal{L}^{\epsilon}_{k})^{N_{k}}(\bar{\mathcal{L}}^{\epsilon}_{k'})^{N_{k'}},\prod_{k,k'}(\mathcal{L}^{\epsilon}_{-k})^{N_{k}}(\bar{\mathcal{L}}^{\epsilon}_{-k'})^{N_{k'}}\right]|0_{\epsilon}\rangle} + \nonumber \\
   &+ \frac{\langle 0_{\epsilon}|O_{L}(z,\Bar{z})\left[\prod_{k,k'}(\mathcal{L}^{\epsilon}_{k})^{N_{k}}(\bar{\mathcal{L}}^{\epsilon}_{k'})^{N_{k'}},O_{L}(0,0)\right]\prod_{k,k'}(\mathcal{L}^{\epsilon}_{-k})^{N_{k}}(\bar{\mathcal{L}}^{\epsilon}_{-k'})^{N_{k'}}|0_{\epsilon}\rangle}{\langle 0_{\epsilon}|\left[\prod_{k,k'}(\mathcal{L}^{\epsilon}_{k})^{N_{k}}(\bar{\mathcal{L}}^{\epsilon}_{k'})^{N_{k'}},\prod_{k,k'}(\mathcal{L}^{\epsilon}_{-k})^{N_{k}}(\bar{\mathcal{L}}^{\epsilon}_{-k'})^{N_{k'}}\right]|0_{\epsilon}\rangle} + \nonumber \\
   &+ \frac{\langle 0_{\epsilon}|O_{L}(z,\Bar{z})O_{L}(0,0)\left[\prod_{k,k'}(\mathcal{L}^{\epsilon}_{k})^{N_{k}}(\bar{\mathcal{L}}^{\epsilon}_{k'})^{N_{k'}},\prod_{k,k'}(\mathcal{L}^{\epsilon}_{-k})^{N_{k}}(\bar{\mathcal{L}}^{\epsilon}_{-k'})^{N_{k'}}\right]|0_{\epsilon}\rangle}{\langle 0_{\epsilon}|\left[\prod_{k,k'}(\mathcal{L}^{\epsilon}_{k})^{N_{k}}(\bar{\mathcal{L}}^{\epsilon}_{k'})^{N_{k'}},\prod_{k,k'}(\mathcal{L}^{\epsilon}_{-k})^{N_{k}}(\bar{\mathcal{L}}^{\epsilon}_{-k'})^{N_{k'}}\right]|0_{\epsilon}\rangle}
\end{align}
Using (\ref{primary1}), we can see the numerators of the first two terms of the above expression is $c_{eff}$-independent. However the denominator of all the terms or the norm is proportional to $\mathcal{O}(c_{eff}^{\#})$(for sufficiently small $\epsilon$), where $\#$ is some positive power depends on the $N_{lev},\Bar{N}_{lev}$. Now if we take $\epsilon \rightarrow 0$ limit, we have $c_{eff} \rightarrow \infty$. For instance, if we compute the norm $\langle 0_{\epsilon}| \mathcal{L}^{\epsilon}_{k}\mathcal{L}^{\epsilon}_{-k}|0_{\epsilon}\rangle = \frac{c_{eff}}{12}(k^{3}+k d).$ In $\epsilon \rightarrow 0$, $k$ becomes continuous and $c_{eff} \rightarrow \infty$. This explains at finite $\epsilon$ the descendant correlator becomes hard to compute. Also as we have explained, the vacuum reduces to vacuum of radial quantization in the same limit. Hence finally we will arrive at
\begin{align}\label{2 point}
   \lim_{\epsilon\rightarrow 0 \; \text{or}, \; c_{eff} \rightarrow \infty}  \frac{{}^{\epsilon}\langle\psi|O_{L}(z,\bar{z})O_{L}(0,0)|\psi\rangle^{\epsilon}}{{}^{\epsilon}\langle\psi|\psi\rangle^{\epsilon}} \approx \langle 0| O_{L}(z,\bar{z}) O_{L}(1,1) |0\rangle
\end{align}
Using (\ref{t,s curve}), the vacuum two point function for any cft yields
\begin{align}\label{vac 2pt}
    \langle 0|O_{\Delta}(t,0)O_{\Delta}(0,0) |0\rangle = \frac{d^{\Delta}}{\sinh^{2\Delta}(\frac{\sqrt{d}}{2}t)}
\end{align}
Hence this is a thermal two point function satisfying KMS condition. This is also true for generalized free fields(GFF) in any state since it factorizes into vacuum two point functions. Thus we have seen the vacuum descendants at $\epsilon \rightarrow 0$ reduces to the thermal answer.

\underline{\textit{Primary two point function:}} In a similar way, it is generically hard problem to compute two point function in the primary state $|h,h\rangle^{\epsilon}$ at finite $\epsilon$ since we do not know the explicit form of the state $|0_{\epsilon}\rangle$ \footnote{Even though in the subsequent part we will argue it to be the conformal boundary state, it is still hard to compute boundary state two point function in any cft.}. From (\ref{in}) and (\ref{out}) in the limit $\epsilon \rightarrow 0$, we have $|h,h\rangle = O_{h,h}(z=z_{+},\Bar{z}=z_{-})|0\rangle$ and $\langle h,h| = \langle 0|O_{h,h}(z=z_{-},\Bar{z}=z_{+})$. Since the vacuum $|0\rangle$ shares the vacuum for radial quantization, we can compute the two point correlator in any primary state via the state-operator correspondence. 
Now depending on the weight of the primaries we will get different correlators e.g. HHLL(heavy-heavy-light-light) or LLLL. We will study them in some limiting case where the dominating conformal block of the four point function is known.
\begin{itemize}
    
\item \textit{HHLL:}
If we write the HHLL correlator in the standard form where the four operators are located at $(0,\infty,z,1)$, the identity block in the t-channel dominates in semiclassical limit($c\rightarrow \infty$) with $h_{L}/c <<1$, $h/c$ small but fixed \cite{Fitzpatrick:2014vua, Fitzpatrick:2015zha}. In this limit for $h(H)>c/24$, it will give precisely the thermal two point function once we map it back to cylinder \cite{Fitzpatrick:2014vua} \footnote{Since in the cylinder the time translation generates the Hamiltonian of radial quantization}. In our case, the map is different from $z \rightarrow \omega$ and the heavy states are not the traditional in and out states defined in the radial quantization. In fact, since  we can write $O_{h,h}(z_{+},z_{-} = e^{z_{+}L_{-1}} e^{z_{-}\bar{L}_{-1}}O_{h,h}(0)|0\rangle$, these are global descendants of standard in/out state in radial quantization. However we will see after doing a suitable global transformation we can exactly get the standard HHLL form which can give the thermal answer for $h(H)<c/24$ in the quantization time $t$. Since in (\ref{2 point}), three points are fixed, we can rewrite that in terms of cross ratio($z',\Bar{z}'$) in the following way
\begin{align}\label{2 point 2}
    &\frac{\langle h,h| O_{L}(z,\bar{z}) O_{L}(0,0) |h,h\rangle}{\langle h,h|h,h\rangle} \nonumber \\
    & = \left(\frac{dz'}{dz}\right)^{h_{L}}\left(\frac{d\bar{z}'}{d\bar{z}}\right)^{h_{L}}(z_{+}-z_{-})^{4h}\langle 0|O_{h,h}(z_{-},z_{+})O_{L}(z'(z),\bar{z}'(\bar{z}))O_{L}(0,0)O_{h,h}(z_{+},z_{-}) |0\rangle
\end{align}
Here
\begin{align}
    z'(z) = \frac{z_{-}(z-z_{+})}{z_{+}(z-z_{-})} , \; \frac{dz'}{dz} = \frac{(z_{+}-z_{-})z_{-}}{(z-z_{-})^{2}z_{+}}
\end{align}
We will now perform the following global conformal transformation to get the four point function in the desired form
\begin{align}
    \zeta = \frac{z_{-}z_{+}}{(1-z_{+})(z-z_{-})}
\end{align}
This is of the same form of $z'$. In doing this transformation, the jacobian coming from $\left(\frac{d\zeta}{dz'}\right)^{h_{L}}$ exactly cancels $\left(\frac{dz'}{dz}\right)^{h_{L}}$. Finally (\ref{2 point 2}) reduces to
\begin{align}\label{4 pt}
   &\frac{\langle h,h| O_{L}(z,\bar{z}) O_{L}(0,0) |h,h\rangle}{\langle h,h|h,h\rangle} \nonumber \\
   &= \left(\frac{|z_{+}-z_{-}|^{2}}{z_{+}z_{-}}\right)^{2h_{L}}\lim_{\zeta',\Bar{\zeta'} \rightarrow \infty}\zeta'^{2h}\Bar{\zeta'}^{2h}\langle 0|O_{h,h}(\zeta',\bar{\zeta'})O_{L}(\zeta,\Bar{\zeta})O_{L}(1,1)O_{h,h}(0,0)|0\rangle
\end{align}
Here we have considered scalar $O_{L}$ for simplicity. The above expression is of the desired form we want. In terms of $\omega=t+is$, from (\ref{t,s curve}) we have
\begin{align}\label{zeta t s map}
    \zeta = \frac{z_{-}}{z_{+}} e^{i\omega\sqrt{d}} = e^{i\sqrt{d}(t'+is)} \; \text{where} \; t' = t+\frac{1}{\sqrt{d}}Arg\left(\frac{z_{-}}{z_{+}}\right)
\end{align}
Similarly we have $\zeta' = e^{-i\sqrt{d}(t'-is)}$. For $\alpha = 0$, the Jacobian pre-factor $\frac{|z_{+}-z_{-}|^{2}}{z_{+}z_{-}}$ becomes 1 and $t'=t+\pi$. Choosing $s=0$ and using the expression for identity block for $h<\frac{c}{24}$ as in \cite{Fitzpatrick:2014vua}, we finally obtain after analytic continuation $t\rightarrow it$ :
\begin{align}\label{HHLL 2pt}
    &\frac{\langle h,h| O_{L}(it) O_{L}(0) |h,h\rangle}{\langle h,h|h,h\rangle} =  \frac{(\pi T)^{2h_{L}}}{\sinh^{2h_{L}}(\pi T t)} \; \text{where} \; T = \frac{\sqrt{d(1-24h/c)}}{2\pi}
\end{align}
This is exactly the thermal two point function\footnote{One could get this by considering cft two point function on a thermal cylinder. Since in this quantization the spatial direction $s$ is not circle due to fixed points, we do not get torus answer.}. This satisfy KMS condition once we analytically continue $it \rightarrow z$ in a finite strip with $0\leq Im(z) \leq \frac{1}{T}$. Note that, for $h>c/24$ we need to change $T \rightarrow iT$ and from (\ref{HHLL 2pt}) we get two point function to be periodic in Lorentzian time. Remarkably, this behavior is exactly opposite to the radial quantization where for primaries with $h>c/24$ the two point function behaves like thermal and for $h<c/24$ the behavior is periodic in time. This fact has important implication in holography and typical state which we will discuss in later sections.

\item \textit{LLLL:} If we consider $h$ to be light operator such that $\frac{h h_{L}}{c}$ is small but fixed, the LLLL identity block $g(\zeta)$ will be of the form \cite{Fitzpatrick:2014vua}
\begin{align}
    g(\zeta) &= \exp\left(2\frac{h h_{L}}{c}\zeta^{2}{}_{2}F_{1}(2,2,4,\zeta)\right) \approx 1+2\frac{h h_{L}}{c}\zeta^{2}{}_{2}F_{1}(2,2,4,\zeta) \nonumber \\
    & = 1+ 2\frac{h h_{L}}{c}\left( \frac{6(\zeta-2)\log(1-\zeta)}{\zeta}-12\right) + \mathcal{O}(\frac{1}{c^{2}})
\end{align}
We can substitute the above in the four point function (\ref{4 pt}) and using (\ref{zeta t s map}) we can get the final answer. One could explicitly check that the expression (after suitable conformal transformation) is decaying in time which makes it thermal. It can be shown to satisfy KMS condition in large $c$. 

\end{itemize}

\underline{\textit{Connection to angular quantization and thermal correlators:}} 

The central idea of angular or Rindler quantization as depicted in \cite{Agia:2022srj}, is that in a sphere conformal frame, the two fixed points(in the Rindler conformal frame which corresponds to the two endpoints of the half sided Rindler region) lies in the two poles and the angular time direction creates states which lives on the meridian connecting two poles. The natural way to find correlators in this setting is by taking thermal traces in the angular direction where periodic boundary conditions on fields imposed by angular time. However due to the presence of the fixed points, the thermal trace is ill defined and hence the Hilbert space is also not well-defined. To get a physical sensible Hilbert space, one must impose a regularization procedure by cutting a hole around fixed points. In this regulated settings, one could in principle compute a thermal trace and thermal correlators which at the end should give the originally desired answer by taking a shrinking limit of the holes to the fixed points. One can also use a conformal map under which the whole set up transforms to a finite cylinder which is topologically similar to an annulus. In the cylinder frame Euclidean time is circular and the space direction has a cut-off($\epsilon$) on both sides. Hence partition function or sphere correlators can be written as $Tr_{\mathcal{H}_{\epsilon}}(e^{-2\pi H_{\epsilon}}O_{1}O_{2}\dots )$. The states of the Hilbert space $\mathcal{H}_{\epsilon}$ must obey boundary condition at two spatial end points of the cylinder.

On the other hand, the same picture can be visualized as time evolution of CFT on a circle with radius $\epsilon$ to a circle of radius $1/\epsilon$ in the path integral picture on annulus by the virtue of open-closed string duality\footnote{This is the stereographic projection of sphere conformal frame where two fixed points are getting mapped to $0$ and $\infty$.}. In this open string picture, the same correlators can be computed as a transition amplitude between two boundary condition $|B_{\epsilon}\rangle$ and $|B_{\frac{1}{\epsilon}}\rangle$ as $\langle B_{\epsilon}| e^{-H_{cft}/W}O_{1}O_{2}\dots | B_{\frac{1}{\epsilon}}\rangle$. Here $H_{cft}$ is the Hamiltonian in radial quantization and $W$ is the width of the annulus. The main proposal of \cite{Agia:2022srj} is to write correlators of angular quantization in the $\epsilon\rightarrow 0$ limit as the following:
\begin{align}
     \lim_{\epsilon \rightarrow 0}Tr_{\mathcal{H}_{\epsilon}}(e^{-\beta H_{\epsilon}}O_{1}O_{2}\dots) \propto \langle \mathcal{O}_{i}|O_{1}O_{2}\dots|\mathcal{O}_{j}\rangle
\end{align}
Here $\mathcal{O}_{i,j}$ are two local primary operators at the fixed points which can be obtained from the shrinking limit of $B_{i,j}$. However which operators are appearing in the final spectrum is fully determined by the choice of boundary conditions we are using to obtain this. In this way the quantization is very much dependent on the boundary conditions on the regulator and it is dubbed as `asymptotics/operator correspondence'\cite{Agia:2022srj}.

This whole story can be generalized to our case in a straightforward way. From (\ref{t,s curve}) one can choose a conformal frame $\zeta \equiv \frac{z-z_{+}}{z-z_{-}} = e^{i\omega\sqrt{d}}$. One can think $\zeta\rightarrow \omega$ is a map from plane to thermal finite (after putting a cut-off) cylinder. In this frame we can similarly define a thermal trace or thermal correlators as we defined previously. The two fixed points are getting mapped to $s=-\infty,\infty$ respectively(in the $\epsilon \rightarrow 0$ limit). Going to sphere conformal frame is also trivial where those are mapped back to 0 and $\infty$. Thereby the rest of the analysis to angular quantization is straightforward to this case where we could cut a hole around a fixed points and define states which has a shrinking limit. Hence the equivalence of thermal two point function and four point sphere correlators in modular quantization in the $\epsilon \rightarrow 0$ limit is guaranteed. This in turn shows that only primaries will be appeared in the $\epsilon \rightarrow 0$ limit and they are placed at the fixed points.  However, in stead of using angular quantization technique, we provide an independent quantization scheme by unwrapping $t$ circles to get the covering space of the quantization space-time. In our case, the Lorentzian representation of the modular quantization still advocates a `state/operator correspondence' unlike the `asymptotics/operators' correspondence as described in \cite{Agia:2022srj}. However the end results are parallel as we see here.

In our setting, the choice of boundary condition plays an important role since we demanded $\lim_{\epsilon \rightarrow 0}|0_{\epsilon}\rangle \rightarrow |0\rangle$. In our quantization scheme, $|0_{\epsilon}\rangle$ must be related to $|B_{\epsilon}\rangle$\footnote{Since the construction of $|0_{\epsilon}\rangle$ is performed(in path integral) at $t\rightarrow -\infty$ where the cut-off is placed.}. Here we propose a way to fix the boundary condition. In quantization coordinate, $s$ generates the flow from one fixed point to the other one. The conserved charge corresponding to the generator of the flow is given by $P_{D} \equiv \mathcal{L}_{0}-\Bar{\mathcal{L}}_{0}$ \footnote{This notation was introduced and used in \cite{Czech:2017zfq, Das:2019iit, Das:2020goe} in the context of modular flow. There $P_{D}$ generates the flow from one endpoint to the other end-point of the interval.}. Now imposing boundary condition at the two cut-off implies demanding $P_{D} = 0$ \footnote{This is analogous to imposing the off-diagonal stress tensor perpendicular to the boundary of the upper half plane to be vanishing or $T_{tx} = 0$.}. This is natural requirement of demanding no momentum flow across the boundary. In a similar fashion $P_{D} = 0$ naturally imposes local boundary conditions on the both cut-off. Hence the natural boundary state in our construction is
\begin{align}
    P_{D}|B\rangle = 0
\end{align}
In terms of generators of radial quantization this implies
\begin{align}
    \Bigg[\alpha(L_{0}-\Bar{L}_{0}) + \beta\sum_{n=-1,1}(L_{n}-\Bar{L}_{-n})\Bigg]|B\rangle = 0
\end{align}
One standard solution to this is the Ishibashi state \cite{Ishibashi:1988kg} which is of the form $\sum_{N=0}^{\infty}\sum_{j=0}^{d(N)}|h,N,j\rangle \otimes |\overline{h,N,j}\rangle$, where $d(N)$ denotes the degeneracy of descendant level $N$. $|h,N,j\rangle$ is orthonormal basis. However since $P_{D}$ is generated  only by global Virasoro descendants, we can have a restricted solution of the form \cite{Castro:2018srf}
\begin{align}
    |h\rangle\rangle = \sum_{n=0}^{\infty}c_{m}^{2} L^{m}_{-1}\Bar{L}^{m}_{-1}\mathcal{O}_{h,h}(z_{+},z_{-})|0\rangle
\end{align}
Here $L_{-1}$ is defined on the cut-off circle of radius $\epsilon$. 
\begin{align}
    L_{-1} \equiv \oint_{z_{+}}dz T(z), \; \bar{L}_{-1} \equiv \oint_{z_{-}}d\bar{z} \bar{T}(\bar{z})
\end{align}
Using change of variable $z-z_{\pm} = \epsilon e^{i\theta_{\epsilon}}$ , we get the integrals are proportional to $\epsilon$. Thus at $\epsilon \rightarrow 0$ the contributions to the descendants are vanishing \footnote{We want to thank Bobby Ezhuthachan to clarify this point.}. However a physical boundary state should be a linear combination of Ishibashi states, $|B\rangle^{\epsilon} = \sum_{h}a_{h}^{\epsilon}|h\rangle\rangle$. By choosing $a_{h}^{\epsilon}$s appropriately one could obtain the desired condition i.e.
\begin{align}\label{rel1}
  \lim_{\epsilon\rightarrow 0}  |B\rangle^{\epsilon} \rightarrow |0\rangle
\end{align}
Since the boundary states are non-normalizable, we should take normalized states of the form $e^{-\epsilon H_{\epsilon}}|B\rangle^{\epsilon}$. In order to make connection with modular quantization, we have 
\begin{align}\label{rel2}
  |0\rangle^{\epsilon} \sim e^{-\epsilon H_{\epsilon}}|B\rangle^{\epsilon}  
\end{align}
To summarize, using (\ref{in}), (\ref{out}), (\ref{2 point}), (\ref{rel1}) and (\ref{rel2}) we propose a slight modification of the claim of angular quantization by \cite{Agia:2022srj} as the following
\begin{align}
    \lim_{\epsilon \rightarrow 0}Tr_{\mathcal{H}_{\epsilon}}(e^{-\beta H_{\epsilon}}O_{1}O_{2}\dots) \propto \sum_{i=|0\rangle, |h,h\rangle}\langle i|O_{1}O_{2}\dots|i\rangle
\end{align}
In the language of `angular quantization' this modification implies sum over all possible boundary condition to get different possible operators(identity or primaries) at the fixed points.
 \section{Entropy and density of states}\label{sec4}
Here we exploit the equivalence of annulus partition function in two different channel to derive both canonical and microcanonical entropy in the quantization. This is motivated from the procedure of angular quantization which we described briefly in the last section. Firstly, we can think the annulus or finite cylinder partition function as the CFT partition function of infinite strip of width $W$ which we identify periodically in $t$ direction. The corresponding Hamiltonian is the spacetime generator of infinitesimal translation along the strip which in our case is just $H_{\epsilon} = \mathcal{L}_{0}^{\epsilon}+\bar{\mathcal{L}}^{\epsilon}_{0}$. In the notation of \cite{Cardy:2016fqc}, the correct partition function can be written as
\begin{align}\label{part1}
    Z_{\epsilon} = Tr e^{-\frac{2\pi^{2}}{W}H_{\epsilon}} =  \int dE \rho(E) e^{-\frac{2\pi^{2}}{W}E}
\end{align}
Here $\rho(E)$ is the density of states of energy $E=\Delta$. The modular parameter of annulus is $q = e^{-\frac{2\pi^{2}}{W}}$. There is another modular parameter $\Tilde{q} = e^{-2W}$. This appears when we do the path integral in other channel where $s$ generates translation between two boundary states. In other words, we can view the same annulus partition function as the transition amplitude between two boundary states where the Hamiltonian is $\Tilde{H} = L_{0}+\Bar{L}_{0}-\frac{c}{12}$.
\begin{align}
    Z_{\epsilon} \approx \langle B_{\epsilon} | e^{-2W\Tilde{H}} |B_{\frac{1}{\epsilon}}\rangle = \sum_{h,\Tilde{h}}|a_{h}^{\epsilon}|^{2} \langle\langle h | \Tilde{h}\rangle\langle\Tilde{h} | h\rangle\rangle e^{-4W\Tilde{\Delta}+\frac{c}{6}W}
\end{align}
In our case $W = \log(\frac{\sqrt{d}}{\beta\epsilon})$. If we take $\epsilon \rightarrow 0$ limit then $W \rightarrow \infty$. In this large width limit the leading contribution of the above $\Tilde{\Delta}$ sum comes from the vacuum $|0\rangle$. This gives
\begin{align}
    Z_{\epsilon\rightarrow 0} \sim \lim_{\epsilon \rightarrow 0}\sum_{h,\Tilde{h}}|a_{h}^{\epsilon}|^{2} \langle\langle h | 0\rangle\langle 0 | h\rangle\rangle e^{\frac{c}{6}W}
\end{align}
Since as we demanded, in $\epsilon \rightarrow 0$, $|B_{\epsilon}\rangle \rightarrow |0\rangle $ the above expression yields
\begin{align}\label{part2}
    Z_{\epsilon \rightarrow 0} \sim e^{\frac{c}{6}W}
\end{align}
In (\ref{part1}), we can treat $\frac{2\pi^{2}}{W}$ as the effective inverse temperature $\beta_{eff}$ in a canonical ensemble. Hence from (\ref{part2}) we can evaluate the thermal entropy corresponding to the ensemble as the following:
\begin{align}\label{canonical ent}
    S_{\epsilon \rightarrow 0} &= \frac{\partial}{\partial(\frac{1}{\beta_{eff}})}\frac{\ln{Z_{\epsilon\rightarrow 0}}}{\beta_{eff}} = \frac{\partial}{\partial W}\left(\frac{c}{6}W^{2}\right) \nonumber \\
    &= \frac{c}{3}\ln{\frac{\sqrt{d}}{\beta\epsilon}} = \frac{c}{3}\ln{\left(\frac{|z_{+}-z_{-}|}{\epsilon}\right)}
\end{align}
This is precisely the entanglement entropy $S_{EE}$ of the interval $(z_{+},z_{-})$. In other words, the thermodynamic entropy corresponding to the canonical ensemble defined in the annulus reproduces the entanglement entropy of the interval in which we quantize the total modular Hamiltonian in $\epsilon \rightarrow 0$ limit. From the bulk side this can be realized as the thermal entropy of AdS-Rindler as described in \cite{Das:2024vqe}.

Also from (\ref{part1}) and (\ref{part2}), we will have an expression for high-energy density of states as the inverse Laplace transform of the Partition function 
\begin{align}
    \rho(E) = \frac{1}{2\pi i}\int^{C+i\infty}_{C-i\infty}dW e^{\frac{c}{6}W+\frac{2\pi^{2}}{W}E}
\end{align}
We can solve the integral by saddle point approximation. The saddle point is $W^{*} = \sqrt{\frac{12E}{c}}\pi$. This in turn fixes $\epsilon$ in terms of $E/c$. More precisely $\epsilon^{*} \sim \mathcal{O}(e^{-\sqrt{\frac{12E}{c}}})$. If $E \sim \mathcal{O}(c)$, $\epsilon^{*}$ becomes $\mathcal{O}(1)$. Hence the asymptotic density of state in the Hilbert space is dominated by the high-energy eigenstates at finite $\epsilon$. One can only take $\epsilon \rightarrow 0$ limit for finite $c<<E$. However to compare with BTZ entropy one must take $E=\frac{c}{12}$ and hence we can not take $\epsilon \rightarrow 0$.  Putting $W^{*}$ back into the integral we finally obtain:
\begin{align}
    \rho(E) = \exp\left(2\pi\sqrt{\frac{cE}{3}}\right)
\end{align}
Remarkably, this expression coincides with the well-known asymptotic Cardy density of states \cite{Cardy:1986ie} in radial quantization. The corresponding microcanonical entropy for a fixed energy $E$ is given by
\begin{align}\label{Microcanonical ent}
    S(E) = \log\rho(E) = 2\pi\sqrt{\frac{cE}{3}}
\end{align}
This is the famous Cardy entropy which in the bulk coincides with the entropy of BTZ blackhole when $E$ is taken to be the mass of the BTZ ($M_{BTZ}\equiv \frac{c}{12}$) \cite{Strominger:1997eq}. The temperature of BTZ also coincides with $\frac{2\pi^{2}}{W^{*}} = 2\pi$ (with $r_{h}=1$) when $E=M_{BTZ}$. Note that, here we implicitly take holographic CFTs to realize the microcanonical entropy as the spherical BTZ entropy. Even though, in both radial and modular quantization the high-energy states reproduce the BTZ entropy, the nature of the eigenstates are significantly different from each other. The high-energy states in  modular quantization which contributes to the BTZ entropy will correspond to high descendant states of $|0\rangle_{\epsilon}$, high energy primaries $|h,h\rangle^{\epsilon}$ with $h =\frac{c}{48\beta\epsilon} > \frac{c}{24}$(for $\sqrt{d}=1$ \footnote{Since here $E$ corresponds to the eigenvalue of the Hamiltonian $H_{\epsilon}$ as in (\ref{new vac}).} at fixed finite $\epsilon$, as we mentioned. Note that, the canonical (\ref{canonical ent}) and microcanonical entropy (\ref{Microcanonical ent}) will only match at the strict semiclassical limit $c \rightarrow \infty$ as both of them becomes diverging.
\section{AdS description}\label{sec5}

Here we discuss the description of bulk AdS$_{3}$ metric dual to different  vacuum of $H$: $|0\rangle$, $|h,h\rangle$ with $h\sim\mathcal{O}(c)$. We will also discuss possible interpretation to finite $\epsilon$ vacuum descendant states in terms of stretched horizon quantization. Most of these discussions are subjected to hold in semiclassical limit of AdS/CFT $(c\rightarrow \infty, G_{N}\rightarrow 0)$ where we have smooth description of metrics.

\underline{\textit{Vacuum}:} Since the vacuum $|0\rangle$ is annihilitated by $L_{\pm 1,0}$, the modular Hamiltonian $H$ does not change the state and hence in the bulk we still have pure AdS$_{3}$. This observation was used in \cite{Das:2022pez} to get the bulk metric by solving bulk curve equation generated by the bulk Hamiltonian corresponding to the boundary one. The bulk metric was observed to be an AdS$_{2}$ blackhole patch (in terms of $t$ and $s$ coordinate of the quantization) in AdS$_{3}$. In our companion paper \cite{Das:2024vqe}, it is pointed out that the fixed points of the quantization will correspond to horizon of AdS$_{2}$ blackhole patch and in the bulk this corresponds to the RT surface. In particular, the 2d boundary AdS$_{2}$ blackhole effectively captures the modular quantization.

However, since the boundary Hamiltonian is the vacuum (total) modular Hamiltonian of a subregion in $S^{1}$, one would still expect the bulk could be described by the corresponding entanglement wedge which will be the same of a AdS-Rindler patch\footnote{We would like to thank Chethan Krishnan for emphasizing this point.}. In vacuum, the AdS$_{3}$ metric in Poincare patch is written as:
\begin{align}\label{Poin}
    ds^{2} = \frac{du^{2}+d\omega d\bar{\omega}}{u^{2}}
\end{align}
An asymptotically AdS$_{3}$ vacuum metric can be written in the Fefferman-Graham gauge as the following:
\begin{align}\label{FG}
    ds^{2} = \frac{dy^{2}+dzd\bar{z}}{y^{2}}+ \frac{L(z)}{2}dz^{2}+\frac{\bar{L}(\bar{z})}{2}d\Bar{z}^{2}+\frac{y^{2}L(z)\Bar{L}(\Bar{z})}{4}dz d\Bar{z}
\end{align}
This class of metrics are also known as Banados metric which is related to pure AdS$_{3}$ by large diffeomorphism \cite{Banados:1994tn}. Here 
\begin{align}\label{Sch}
    L(z) = -\frac{12}{c}T(z) \equiv -\frac{f'''(z)f'(z)-\frac{3}{2}(f''(z))^{2}}{(f'(z))^{2}}
\end{align}
In the boundary $(\omega,\Bar{\omega}) \rightarrow (f(z),\bar{f}(\Bar{z}))$ is local conformal transformation which generates a (generically) non-vanishing stress tensor which is the source of the metric (\ref{FG}). The explicit form of the bulk diffeomorphism from $(u,\omega,\Bar{\omega}) \rightarrow (y,z,\Bar{z})$ is also known \cite{Roberts:2012aq} :
\begin{align}\label{diffeo}
    &\omega = f(z) - \frac{2y^{2}(f'(z))^{2}\Bar{f}''(\Bar{z})}{4f'(z)\bar{f}'(\bar{z})+y^{2}f''(z)\Bar{f}''(\Bar{z})} \\
    &\bar{\omega} = \bar{f}(\bar{z}) - \frac{2y^{2}(\bar{f}'(\bar{z}))^{2} f''(z)}{4f'(z)\bar{f}'(\bar{z})+y^{2}f''(z)\Bar{f}''(\Bar{z})} \\
    & u = y\frac{4(f'(z)\bar{f}'(\Bar{z}))^{3/2}}{4f'(z)\bar{f}'(\bar{z})+y^{2}f''(z)\Bar{f}''(\Bar{z})}
\end{align}
Here we want to find the bulk metric corresponding to the boundary transformation (\ref{t,s curve}) which basically maps complex plane to the new cylindrical geometry generated by $(t,s)$.
To begin with, we use the conformal transformation $\zeta = \frac{\omega-z_{+}}{\omega-z_{-}}$. This is a global conformal transformation and hence $L=\Bar{L} = 0$. We can find a new coordinates $(u,\omega,\Bar{\omega}) \rightarrow (u',\zeta,\Bar{\zeta})$ for which the metric remains Poincare. From (\ref{FG}), we get
\begin{align}
    ds^{2} = \frac{du'^{2}+d\zeta d\Bar{\zeta}}{u'^{2}}
\end{align}
In this frame, as we mentioned the fixed points $(z_{+},z_{-}) \rightarrow (0,\infty)$. In the frame of modular time, the coordinate transformation is determined by $\zeta = e^{iz\sqrt{d}}$, where $z=t+is$. Hence the Banados metric under this transformation is sourced by the stress tensor $L=\Bar{L}=-\frac{d}{2}$. Using (\ref{FG}), the corresponding metric has the following form:
\begin{align}\label{vac metric1}
    ds^{2} = \frac{dy^{2}+dzd\bar{z}}{y^{2}}- \frac{d}{4}dz^{2}-\frac{d}{4}d\Bar{z}^{2}+\frac{y^{2}d^{2}}{16}dz d\Bar{z}
\end{align}
After using the following transformation $z=t+is,\bar{z}=t-is$ and $r^{2} = \frac{(dy^{2}+4)^{2}}{16y^{2}}$ the final form of the metric is
\begin{align}\label{ads rind}
    ds^{2} = (r^{2}-d)dt^{2}+\frac{dr^{2}}{r^{2}-d}+r^{2}ds^{2}
\end{align}
This is the metric for AdS-Rindler patch or (planar) BTZ \footnote{In the next section we will argue that both gives the same description in $G_{N} \rightarrow 0$ limit.} with the horizon radius $r_{h}=\sqrt{d}$. Here $r:(\sqrt{d},\infty)$. Since the vacuum (of radial quantization) remains invariant under the action of modular Hamiltonian, the pure AdS$_{3}$ will not change. Hence from an observer point of view, this is an observer horizon. Now at $r=\sqrt{d}$ one can check that $y^{2} = \frac{4}{d}$. On the other hand, using (\ref{diffeo}) we have
\begin{align}
    \zeta = \frac{4-y^{2}d}{4+y^{2}d} e^{iz\sqrt{d}}, \; \bar{\zeta} = \frac{4-y^{2}d}{4+y^{2}d} e^{i\bar{z}\sqrt{d}}, \; u' = \frac{4y\sqrt{d}}{4+y^{2}d}
\end{align}
From the above relations, we can easily see that at $y^{2} = \frac{4}{d}$, $\zeta=\Bar{\zeta} = 0$. Thus $r=\sqrt{d}$ corresponds to one of the fixed points $\zeta = 0$. The same horizon can be re-written in terms of the other fixed point as the following. If we use the map $\zeta' = \frac{\omega-z_{-}}{\omega-z_{+}}$ and use the map $\zeta' = e^{-iz\sqrt{d}}$ we will end up again with the same metric. The only difference will be that the other fixed point $z=z_{-}$ is dual to the horizon. The map suggests a change from $z \rightarrow -z$ to see another fixed point. In Lorentzian picture, $z\rightarrow -z$ implies $t \rightarrow -t$ and $s\rightarrow -s$. This indicates, changing the time direction (backward) we may find another fixed point ($z_{-}$) which corresponds to the horizon. This refers, we may think of one fixed point to be the future horizon and the other to be the past horizon of the AdS-Rindler patch. The metric is also consistent with the boundary limit of scalar two point function as we get in (\ref{vac 2pt}). By analytic continuation of $t \rightarrow it$ in (\ref{ads rind}), we would end up with the standard Lorentzian AdS$_{3}$-Rindler.

\underline{\textit{Primary states or other vacuum:}}
The other class of vacuum of $H$ is denoted as $|h,h\rangle$ which is made out of scalar primaries located at fixed points. We are interested in finding the dual metric when the operator dimension $h \sim \mathcal{O}(c)$ for which we obtained two point functions. A heavy operator with dimension $h \sim \mathcal{O}(c)$ creates a stress tensor which is the source of the metric in $(y,z,\Bar{z})$ coordinates \cite{Fitzpatrick:2015zha, Asplund:2014coa}. In particular, the heavy operators at $(\zeta=0,\infty)$ creates the state $|h,h\rangle$. Hence $\langle h,h| T(z)|h,h\rangle = \frac{h}{z^{2}}$. We can also think of this as the energy sourced by some diffeomorphism $\zeta \rightarrow f(z)$. Thus we can always write
\begin{align}\label{T heavy}
    &{}_{h}\langle T(z) \rangle_{h} = \frac{c}{24z^{2}} - \frac{c}{12}L(z) \\
    &\implies L(z) = \frac{1}{2z^{2}}-\frac{12}{c}\frac{h}{z^{2}} = \frac{1}{2z^{2}}\left(1-\frac{24h}{c}\right) 
\end{align}
The first term of (\ref{T heavy}) comes from the Schwarzian term from global($w$) to Poincare map ($w \rightarrow \log z$) while the second term corresponds to Schwarzian of the map from $\zeta \rightarrow z$. One can check that if $f(z) =z^{\alpha}$, the form of $L(z)$ as in (\ref{T heavy}) is recovered. Hence, by putting $f(z) = e^{iz\sqrt{d}\alpha} $ with $\alpha = \sqrt{1-\frac{24h}{c}}$, then from (\ref{Sch}) $L=-\frac{d\alpha^{2}}{2}$. Hence the metric (\ref{FG}) becomes
\begin{align}\label{vac metric2}
    ds^{2} = \frac{dy^{2}+dzd\bar{z}}{y^{2}}- \frac{d \alpha^{2}}{4}dz^{2}-\frac{d \alpha^{2}}{4}d\Bar{z}^{2}+\frac{y^{2}d^{2}\alpha^{4}}{16}dz d\Bar{z}
\end{align}
In a similar way to the previous vacuum case, we can define $r^{2} = \frac{(d\alpha^{2}y^{2}+4)^{2}}{16y^{2}}$ under which the metric takes the form:
\begin{align}
     ds^{2} = (r^{2}-d\alpha^{2})dt^{2}+\frac{dr^{2}}{r^{2}-d\alpha^{2}}+r^{2}ds^{2}
\end{align}
This is a metric of a planar BTZ for $h<c/24$. In this metric, the horizon is at $r=\sqrt{d}\alpha$ for which $y^{2}=\frac{4}{d\alpha^{2}}$. This also describes an AdS-Rindler patch for which $r:(\sqrt{d}\alpha,\infty)$. Similarly the diffeomorphism map for this case is simply the following:
\begin{align}
     \zeta = \frac{4-y^{2}d\alpha^{2}}{4+y^{2}d\alpha^{2}} e^{iz\sqrt{d}\alpha}, \; \bar{\zeta} = \frac{4-y^{2}d\alpha^{2}}{4+y^{2}d\alpha^{2}} e^{i\bar{z}\sqrt{d}\alpha}, \; u' = \frac{4y\sqrt{d}\alpha}{4+y^{2}d\alpha^{2}}
\end{align}
Once again at the horizon $y^{2}=\frac{4}{d\alpha^{2}}$, $\zeta=\zeta' = 0$ which is the fixed point. In this case, there will be energy peaks at the fixed point since ${}_{h}\langle T(\zeta) \rangle_{h} = \frac{h}{\zeta^{2}}$. 

For $h>c/24$ one could similarly obtain the following metric:

\begin{align}
    ds^{2} = (r^{2}+d\alpha^{2})dt^{2}+\frac{dr^{2}}{r^{2}+d\alpha^{2}}+r^{2}ds^{2}
\end{align}
Here $\alpha = (\frac{24h}{c}-1)$. This is a conical defect geometry where the conical singularity is placed at the centre of the global AdS. Remarkably, this is exactly opposite to the case in radial quantization which we mentioned in two point function computation. In radial quantization, $h>c/24$ primaries are interpreted as btz microstates whereas $h<c/24$ corresponds to horizonless conical defect geometries. We will discuss some aspects of this interchanging behavior in these two different quantization in the next section.


\underline{\textit{Descendant of vacuum and finite $\epsilon$ effect:}}

So far we have only considered the dual description of eigenstates of modular Hamiltonian at $\epsilon =0$ which eventually reduces to different vacuum in the same limit. However from the previous section, we have seen the  finite $\epsilon$ states with energy $E=\frac{c}{12}$ would describe blackhole microstates. The important observation we made is the identification of fixed point with the horizon of AdS-Rindler. Hence it is natural to think that cut-off around fixed point will correspond to a stretched horizon which can be realized both in Euclidean and Lorentzian bulk picture. Since our quantization used Lorentzian representation by unwrapping $t$ from one fixed point to the other, we may think the stretched horizon is also stretched over all time from future horizon to past horizon. Thus the natural identification of the cut-off in the bulk is a boundary condition placed at $r=r_{h}+\epsilon$. 

In the modular quantization, the cut-off $\epsilon$ also truncates the range of $s$ coordinate. As we mentioned, in the other representation of the vacuum as AdS$_{2}$ blackhole foliation of AdS$_{3}$, the $s$ coordinate is realized as (upto a coordinate transformation) the radial direction of AdS$_{2}$ blackhole \cite{Das:2022pez}. Thus the AdS$_{2}$ stretched horizon can be exactly realized as the quantization cut-off around fixed points. Moreover, the discreteness condition on $k$ as in (\ref{quantization}) is exactly same as the normal modes of a scalar field in the AdS$_{2}$ blackhole with Dirichlet boundary condition on the stretched horizon \cite{Soni:2023fke}. From these facts, one might be able to realize modular quantization as a free scalar field quantization in the stretched horizon background of AdS$_{2}$ blackhole slice. In particular, the (vacuum) descendants of the modular quantization can be recast as excited states on top of the Boulware state in AdS$_{2}$ blackhole.


However here we are looking for a similar realization in the dual AdS-Rindler or planar BTZ metric with stretched horizon. In our recent paper \cite{Burman:2023kko}, we studied free massless scalar field in BTZ stretched horizon (Planckian) background with Dirichlet boundary condition. Even though it was done for BTZ with compactified spatial boundary, the similar analysis will also hold for the BTZ with planar boundary. Following \cite{Burman:2023kko} We can write a massless scalar field $\phi(r,t,s)$ in Lorentzian planar BTZ stretched horizon background as the following\footnote{Here we denote AdS-Rindler metric (\ref{ads rind}) as the planar BTZ.}:
\begin{align}
    \phi(r,t,s) = \int dK\sum_{n}e^{-i\omega_{n} t}e^{iKs}\phi_{n,K}(r)
\end{align}
The Dirichlet boundary condition $\phi(r=\sqrt{d}) = 0$ would give a discrete spectrum ($n \in \mathbb{Z}$) of normal modes
\begin{align}
    \omega_{n} \approx \frac{2n\pi\sqrt{d}}{\log(\frac{\sqrt{d}}{2\epsilon})}
\end{align}
This is again of the same form of (\ref{quantization}) when $n \in \mathbb{Z}$. However there are two subtle difference with the bulk story we presented in \cite{Burman:2023kko}. We considered a Planckian stretched horizon by placing it at a Planck geodesic distance from the Horizon. This further fixed $\epsilon \sim \frac{1}{c^{2}}$. Apart from this, a momentum cut-off $K_{cut}$ will be necessary to get a finite partition function of scalar fields. We fixed $K_{cut}$ by demanding the microcanonical entropy of scalar fields at energy $E=c/12$ to be the black hole entropy. From the boundary quantization, we do not have such fixing except in the microcanonical ensemble, where we need to fix $\epsilon \sim \mathcal{O}(1)$ to reproduce BTZ entropy for $E=c/12$. This apparent mismatch could be resolved if we do not fix the position of stretched horizon in the bulk. In fact, our bulk analysis \cite{Burman:2023kko} would still reproduce the same result of BTZ entropy, Hawking temperature and Hartle Hawking correlator once we use the following condition on $\epsilon$, $K_{cut}$ without explicitly fixing them:
\begin{align}
G_{N}K_{cut}\log\left(\frac{\sqrt{d}}{2\epsilon}\right) = 3\pi\sqrt{d}
\end{align}
This condition is true for even finite $\mathcal{O}(1)$ value of $\epsilon$. The appearance of $K_{cut}$ will be visible in cft analysis once we quantize the CFT in momentum space. Just like we have a cut-off in $s$, we could have a $K_{cut}$ in the momentum space. 

Now in the bulk, we considered probe scalar field two point function in a high energy excited state built in the scalar field quantization. The energy condition is given as
\begin{align}\label{energy cond}
    \int^{K_{cut}}_{-K_{cut}}dK\sum_{n} \omega_{n}N_{n} = E = M_{BTZ}
\end{align}
Here $N_{n}$ is the occupation number. Remarkably, this is quite similar to the level of a generic descendant state as in (\ref{descendant level}). From the microcanonical entropy analysis we must have $N_{lev} = \frac{c}{12} = M_{BTZ}$ once we compare these states as BTZ microstates. Thus apart from the degenracy factor $2K_{cut}$ in (\ref{energy cond}) both looks same. Again we do not expect both to be exactly similar, since as we mentioned, the bulk dual to the modular quantization do have an extra bulk direction where the cut-off is placed. However the similarity of the spectrum of descendant level and bulk energy level is encouraging to us. 

The most important result in \cite{Burman:2023kko} is the emergence of thermal boundary two point correlator in $\epsilon \rightarrow 0$ limit (even when $\epsilon$ or $K_{cut}$ are not fixed individually). In the boundary, we have also computed two point correlation function of scalar primaries in the descendant states at $\epsilon \rightarrow 0$ limit which reproduce the exact thermal two point function (\ref{vac 2pt}) with inverse temperature $\frac{2\pi}{\sqrt{d}}$. This is also the same boundary correlator in Planar BTZ with the same Hawking temperature. Interestingly, if we compute finite $\epsilon$ correlator with $c\rightarrow \infty$, we do not have the same answer. In that limit (\ref{desc corr}) becomes 
\begin{align}
    \lim_{c\rightarrow \infty \; \text{or}, \; c_{eff} \rightarrow \infty}  \frac{{}^{\epsilon}\langle\psi|O_{L}(z,\bar{z})O_{L}(0,0)|\psi\rangle^{\epsilon}}{{}^{\epsilon}\langle\psi|\psi\rangle^{\epsilon}} \approx \langle 0_{\epsilon}| O_{L}(z,\bar{z}) O_{L}(1,1) |0_{\epsilon}\rangle
\end{align}
We do expect this to be non-decaying or non-thermal. From the bulk side, this implies $K_{cut} \rightarrow \infty$ and $\omega_{n}$ remains discrete. The discrete spectrum will prevent the correlators from decaying and we expect a quasi-periodic nature. Since the usual AdS/CFT rules imply that in large $c$ or $G_{N} \rightarrow 0$ limit, all CFT states could be described by metric solution of Einstein gravity, one could naively expect the finite-$\epsilon$ states might have dual Einstein gravity description in the bulk. In principle, this should always be possible. However, we have implicitly mentioned that the regulated Hilbert space $\mathcal{H}_{\epsilon}$ is much bigger than the actual Hilbert space in $\epsilon \rightarrow 0$. Thus by construction, it is more convenient to think the states in $\epsilon \rightarrow 0$ will only have semiclassical bulk description in terms of Einstein gravity.

From all of these observations and facts we would like to conjecture the following:
\begin{itemize}
    \item The (planar) BTZ stretched horizon vacuum in the free scalar field quantization is proportional to $|0_{\epsilon}\rangle$ in the dual CFT.
    \item Excited states on the top of the vacuum in the stretched horizon may have an one-to-one correspondence to generic (vacuum) descendant states in boundary modular quantization as in (\ref{descendant level}).
    \item Signature of smooth horizon or Hartle-Hawking correlators is only emerging in the strict $\epsilon \rightarrow 0$ limit. For finite $\epsilon$ and $G_{N} \rightarrow 0$, we still do not have a signature of smoothness.
    \item Vacuum descendants in the regulated modular quantization in any generic CFT (irrespective of any $c$) has a dual description in terms of EFT of free scalar fields in the (planar) BTZ stretched horizon background which provide an effective field theory description of black hole microstates. 
    \item Bulk effective field theory arises when we take both $\epsilon \rightarrow 0$ and $c\rightarrow \infty$ limit.

\end{itemize}

\section{Discussion}\label{sec6}

In this note, we have realized the appearance and necessity of stretched horizon from conformal field theory in the  AdS/CFT framework. The key point is that we have used a different quantization scheme in CFT, where the construction of a well-defined physical CFT Hilbert space and finite partition function, necessarily leads to the emergence of a stretched horizon in it's holographic dual. 
There is no doubt about the success of radial quantization of CFT to understand bulk physics and bulk thermodynamics in AdS/CFT. However in radial quantization, we do not have a direct encounter with thermal physics or direct appearance of thermal Euclidean time circle, which in the bulk is important to understand near horizon physics. Here we quantize a modular Hamiltonian in which we have a direct realization of the things we just said. Thus this alternate quantization provides a direct understanding of black hole or Rindler physics from where we could get a closer look to near horizon physics. Moreover, we have argued that the high energy cut-off states in the regulated Hilbert space will exactly reproduce Bekenstein-Hawking entropy of BTZ in the microcanonical ensemble. This makes a concrete manifestation of blackhole microstates having no smooth horizon. We will discuss more about our findings and implications in quantum gravity. We begin with summarizing our results.

 \underline{\textit{Summary:}} 
 Here we summarize the key findings of the modular quantization and the possible interpretation of the bulk dual.
 \begin{itemize}
 \item We consider a CFT on spatial ring ($S^{1}$) and quantize it with a sl(2,$\mathbb{R}$) deformed Hamiltonian $H$, which eventually coincides with the total modular Hamiltonian in a sub-segment of the ring. Hence the starting point of modular quantization and radial quantization is same where the difference lies in the form of the Hamiltonian. 
 
     \item The appearance of the fixed points in the quantization is an obstacle to define a notion of thermal trace in the Hilbert space. Also it has an implicit effect to get a continuous and non-normalizable spectrum of the Hilbert space. Once we put a cut-off at $\epsilon$ distance around the fixed points, we get a discrete spectrum, a discrete Virasoro algebra with finite central charge and a finite partition function obtaining by thermal trace over Hilbert space of states. 

     \item To get a well-defined highest weight representation of the Virasoro algebra we unwrapped the non-contractable $t$ circles and stretched it from $-\infty$ to $\infty$. We noted that this procedure constitutes a highest-weight Lorentzian representation of the Euclidean quantization. Those two infinities correspond to the location of two cut-off around fixed points. 

     \item We construct a state $|0_{\epsilon}\rangle$ which plays the similar role to the vacuum in this quantization. We have fixed the limiting behavior of the state in $\epsilon \rightarrow 0$ as $\lim_{\epsilon \rightarrow 0}|0_{\epsilon}\rangle \rightarrow |0\rangle$. We argued from the angular quantization perspective that $|0_{\epsilon}\rangle^{\epsilon} \sim |B\rangle^{\epsilon}$, a linear combination of global Ishibashi states. It also satisfies $\mathcal{L}^{\epsilon}_{k}|0_{\epsilon}\rangle = 0$ for $k\geq 0$

     \item We also construct a class of eigenstate $|h,h\rangle^{\epsilon} \equiv \mathcal{O}_{h,h}(z_{+},z_{-})|0_{\epsilon}\rangle$ which plays the role of the primary states in this quantization satisfying $\mathcal{L}^{\epsilon}_{k}|h,h\rangle^{\epsilon} = 0$ for $k>0$. In $\epsilon \rightarrow 0$ limit, $|h,h\rangle^{\epsilon} \rightarrow |h,h\rangle$. $|h,h\rangle$ forms a class of vacuum for $H$.

     \item Descendants of this quantization are denoted as $(\mathcal{L}^{\epsilon}_{-k_{1}})^{n_{1}} \dots |h,h\rangle^{\epsilon}$. This descendants have finite actions only on $|0_{\epsilon}\rangle$.

     \item To compute two point function in this quantization we first evaluate them in states of finite $\epsilon$ Hilbert space $\mathcal{H}_{\epsilon}$. Later we take $\epsilon \rightarrow 0$ limit. We can compute two point function in generic (vacuum) descendants in $\mathcal{H}_{\epsilon}$. In the $\epsilon \rightarrow 0$ limit, we have seen that those two point function reproduce thermal two point function which is the same as vacuum two point function $\langle 0|O(t)O(0)|0\rangle$. 

     \item  We also obtained $\langle h,h|O(t)O(0)|h,h\rangle$ with $h\sim \mathcal{O}(c)<\frac{c}{24}$ is also thermal with temperature $\frac{\sqrt{d(1-24h/c)}}{2\pi}$ in $c\rightarrow \infty$. For $h \sim \mathcal{O}(1)$ or $h > c/24$, the two point function would be periodic in real time. Remarkably, this temporal behavior contrasts to the two point function in heavy states of radial quantization.

     \item From the consistency condition of annulus partition function, we have computed entropy of the quantization. In particular, we started with a partition function for a canonical ensemble with a fixed temperature defined by the width of the annulus. At $\epsilon \rightarrow 0$, the corresponding thermal entropy we obtained, is the entanglement entropy of the boundary subregion. On the other hand, from the consistency condition of the partition function on annulus, we also obtained an expression for density of state at very high energy. The corresponding microcanonical entropy for high energy states gives the Cardy entropy and for a specific choice of energy $E=c/12$ this deduces to BTZ entropy. This also fixes the temperature to be the Hawking temperature with $\epsilon \sim \mathcal{O}(1)$. The two entropy would be diverging in the strict thermodynamic limit when $c \rightarrow \infty$.

     \item  In the dual AdS$_{3}$ the vacuum $|0\rangle$ is described by AdS-Rindler geometry where the fixed point of the modular quantization corresponds to horizon or RT surface of the geometry. Equivalently, the same geometry also describes a planar BTZ\footnote{Alternatively, this a hyperbolic blackhole which is equivalent to AdS-Rindler in any dimension \cite{Casini:2011kv, Czech:2012be}.}. 
     Since the same quantization describes entanglement entropy or thermal AdS-Rindler entropy in canonical picture and BTZ entropy in microcanonical picture, the vacuum describes AdS-Rindler or planar BTZ in canonical ensemble in the strict $\epsilon \rightarrow 0$ limit. The same vacuum can be described as a TFD state \cite{Czech:2012be} \footnote{This picture is described in detail in \cite{Das:2024vqe}.}. We expect, the microcanonical version of TFD state would describe the BTZ at finite $\epsilon$ \cite{Chandrasekaran:2022eqq}. 

\item The fixed points of the quantization can be identified with horizon (or RT surface) of the AdS-Rindler. Hence the cut-off around the fixed points would play the role of a stretched horizon located outside the horizon.
     \item The other vacuum of the quantization $|h,h\rangle$ for $h \sim \mathcal{O}(c)$ was described by either (planar) BTZ  or conical singularity type geoemtry. In particular, $h<c/24$ describes a BTZ type geometry whereas $h>c/24$ would describe AdS spacetimes with conical singularity. Again this is exactly opposite to the AdS/CFT in the radial quantization.
     \item We have also conjectured the (vacuum) descendants of the modular quantization could be described by excited states of free scalar field theory in the (planar) BTZ background with a dirichlet boundary condition at the stretched horizon. In $\epsilon \rightarrow 0$, they will reproduce thermal two point function in the boundary limit  \cite{Burman:2023kko}.
     \item As we argued, the microstates of BTZ must be eigenstates of $H$ with finite $\epsilon$ at energy $E=c/12$. These could be either vacuum descendants with energy $E=c/12$ or some primaries $|h,h\rangle^{\epsilon}$ with $h \sim \mathcal{O}(c)$. In fact these states do not have a $\epsilon \rightarrow 0$ limit. 
 \end{itemize}

 \underline{\textit{Comments and future direction:}}

 \begin{itemize}
 \item 
 \underline{Emergence of stretched horizon:} The central theme of our paper is to show an emergence of stretched horizon description from CFT under modular quantization. In the usual AdS/CFT description with radially quantized CFT, this emergence is not quite obvious. For instance, in the general bulk prescription of computing entanglement entropy in the bulk using HRRT surface \cite{Ryu:2006bv, Hubeny:2007xt}, a near boundary IR cut-off(without a specific choice of boundary condition) will introduce to regulate divergence. However, our proposal is to introduce a cut-off all along the physical HRRT surface(here the horizon) and impose certain boundary condition to get the same leading order answer\footnote{The author would like to thank the referee of JHEP to point out this distinction}. We would like to justify our proposal in threefold way:
 \begin{itemize}
      \item The primary reason of such a near horizon cut-off originates from the bulk identification of boundary fixed points. We have clearly shown that when we lift the boundary curves generated by modular Hamiltonian into the bulk using the usual AdS/CFT rules, the fixed points are getting mapped to the horizon or the RT surface in AdS-Rindler. Correspondingly the boundary subregion is dual to the entire entanglement wedge or the exterior of the horizon. From this identification, we proposed that cut-off around fixed points will be related to a stretched horizon cut-off just outside the RT surface. In the usual bulk entanglement entropy computation using HRRT surface, this fact will remain obscure. Since in the usual radial quantization, one can not have any fixed points and one does not need the details of modular quantization nor the boundary condition imposed \cite{Ohmori:2014eia}. The fact that we are quantizing modular Hamiltonian, plays the key difference where the stretched horizon cut-off is introduced in the bulk.

    \item In \cite{Das:2024vqe}, we have explained this emergence of near horizon cut-off in a complimentary way. We have first shown that if we lift the boundary modular Hamiltonian into the bulk, the bulk fixed points will give the equation of the RT surface of the corresponding bulk geometry. In this way we can identify the RT surface to the bulk fixed point. Also we have matched the emergent single copy of near horizon Virasoro algebra with central extension term to the Virasoro algebra we get after quantizing the modular Hamiltonian in the presence of cut-off. In doing so, we have to introduce a bulk near horizon cut-off $\epsilon$ which is proportionally related to the near boundary cut-off around the fixed point. This matching explicitly shows that near boundary IR cut-off is related to near horizon UV cut-off. 

    \item In the modular quantization, the leading contribution of the thermal entropy obtained by putting boundary cut-off, precisely matches with the entanglement entropy of the interval as we have shown in section \ref{sec4}. Hence the standard bulk prescription of computing entanglement entropy using RT prescription (by putting near boundary IR cut-off) can be replaced by evaluating thermal entropy of AdS-Rindler. These are identical in the bulk, since in AdS-Rindler, the RT surface coincides with the Rindler horizon. In \cite{Das:2024vqe}, it has been matched exactly. Alternatively, the bulk thermal entropy in AdS-Rindler background could be obtained by the standard procedure of Susskind-Uglum \cite{Susskind:1994sm}, where a stretched horizon cut-off just outside the horizon could be imposed to regulate the divergences of field modes.
 \end{itemize}
 \item \underline{\textit{Algebra of operators: (Type III from Type I)}} Understanding subregions and associated algebra in quantum gravity is still an open and active field of research. Even though, in the presence of spacetime fluctuation, a subregion is ill-defined in perturbative quantum gravity, in the strict $G_{N}\rightarrow 0$ limit an approximated subregion in curved spacetime could be defined. In local QFT, the algebra of observables of a subregion will be a type-III$_{1}$ factor \cite{Witten:2018zxz}. Hence in our settings, one will always get a type-III$_{1}$ von Neumann factor associated to the Hilbert space of the modular quantization\footnote{A more concrete discussion of the algebra type in this quantization can be found in \cite{Das:2024vqe}}. And this is true for all $c$. In the bulk dual, the algebra of quantum fields in AdS-Rindler patch will also constitute type-III$_{1}$ algebra in the strict semiclassical limit $G_{N}\rightarrow 0$ \cite{Bahiru:2022mwh, Leutheusser:2022bgi} \footnote{Note that, here we are considering only algebras of local quantum fields in the fixed back ground. We are neither incorporating the gravitational constraints on the bulk subregions \cite{Jensen:2023yxy} nor considering $1/c$ corrections to fields \cite{AliAhmad:2023etg} for which the type of the algebra could be changed subsequently. However at finite $c$ we would again expect the algebra to be type III$_{1}$. A similar change of algebra type in black hole background has been studied extensively in  \cite{Witten:2021unn,Chandrasekaran:2022eqq,Kudler-Flam:2023qfl}}.
 However there is a puzzling fact about the algebra of observables in black holes. In \cite{Leutheusser:2021qhd, Leutheusser:2021frk}, it is observed that the algebra of local observables for AdS blackholes as described by finite temperature TFD state in radial quantization has an emergent type-III$_{1}$ factor in the strict $G_{N} \rightarrow 0$ limit. This emergence of type III$_{1}$ factor is puzzling only when the boundary manifold is compact. In other words, it was shown the large $N$ single trace sector of the CFT will form this emergent algebra. Thus the apparent puzzle is just an artifact of large $N$ CFT where spectrum becomes continuous. For planar BTZ, this is not the case since the boundary manifold is infinite line for which the algebra will always be type III \cite{Furuya:2023fei}. This is also consistent with our quantization, since the same vacuum could be interpreted as (planar) BTZ in the canonical ensemble in $\epsilon \rightarrow 0$ limit. However, from the bulk perspective, we would expect an emergent notion of type III even for the planar BTZ. Since if we quantize the fields in planar BTZ background we still have a continuous spectrum of quasi normal modes ($\omega$) which has nothing to do with compact or non-compact boundary manifold \cite{Papadodimas:2012aq, Festuccia:2005pi, Festuccia:2006sa}. This continuous spectrum is purely a large $N$ phenomenon and this provides a notion of emergent type III algebra in the bulk. From the boundary side it seems to be contradictory. However, this apparent puzzle could be solved once we identify the boundary Hilbert space as the limiting case of regulated Hilbert space or, $\mathcal{H}_{\epsilon \rightarrow 0} \rightarrow \mathcal{H}$. Also we mentioned the microstates corresponding to the BTZ are those states with finite $\epsilon$ in $\mathcal{H}_{\epsilon}$. The algebra of observables in $\mathcal{H}_{\epsilon}$ should be of type I because of the discrete spectrum. This is also observed in our recent paper \cite{Burman:2023kko} that the bulk fields in the presence of stretched horizon will have discrete normal modes $\omega_{n}$. All these facts suggest the emergence of type III algebra for planar BTZ is related to the emergence of $\epsilon \rightarrow 0$ limit. We should stress that, from our analysis this is not a type-I approximation of type III \cite{Soni:2023fke}, rather we should view this as an emergence of type III from type I in this limit. The same limit could be re-interpreted as $c_{eff} \rightarrow \infty$ or $G_{N,eff} \approx \frac{G_{N}}{\log(1/\epsilon)} \rightarrow 0$.
 The appearance of $c_{eff}$ or $G_{N,eff}$ in AdS/CFT is still not well-understood to us. We hope to get back into this in recent future.

\item  \underline{\textit{Typicality and scars:}} Since the modular quantization is unitarily inequivalent to radial quantization, the status of typical states should also be different in each quantization. The notion of typical state comes from Eigenstate Thermalization Hypothesis(ETH) which motivates why high energy eigenstates of a chaotic Hamiltonian behaves like a thermal state upto exponential correction in entropy. A remarkable feature in radial quantization is that for a sufficiently heavy primary state ($h >\frac{c}{24}$), the heavy-heavy-light-light(HHLL) four point function reproduce the exact thermal two point correlator in the semiclassical limit($c \rightarrow \infty$) \cite{Fitzpatrick:2014vua, Fitzpatrick:2015zha} and thus providing a notion of ETH in those states \cite{Brehm:2018ipf} \footnote{See also \cite{Basu:2017kzo}, where ETH in 2d CFT has been tested for quasiprimary probes in semiclassical limit.}. According to Cardy density of states, these high lying states should correspond to black hole microstates. However this observation will lead us to two puzzling questions:

(i) One can not recover this notion of thermality in the semiclassical limit using high-lying (global)descendant states on some heavy primary($h \sim \mathcal{O}(c) < c/24$) where the total (primary+descendants) scaling dimension of the operators would be $h \geq c/24$. Of course, the technical reason behind this is the structural difference of HHLL vacuum block at large $c$ below and above the threshold of $h = c/24$. The puzzle is why this kind of heavy eigenstates  do not show up as a candidate of typical microstates that ETH predicts\footnote{In \cite{Datta:2019jeo}, a notion of typical CFT microstates build by descendant states at finite $c$ has been discussed. However the notion of thermality is only observed for conserved currents correlators for those states and hence this provides a check for `global ETH'\cite{Basu:2017kzo}. At large $c$, those descendants become primaries and hence for generic operators the notion of thermality will be observed again only for $h>c/24$.}.

(ii) In \cite{Fitzpatrick:2016ive}, the authors raised another puzzle regarding the emergence of periodic singularity in imaginary time in the HHLL correlator for $h>c/24$. In radial quantization, there will be none but ope singulairities in pure state four point function which is guaranteed from unitarity. The authors rephrased this puzzling fact as the information paradox in microcanonical ensemble of pure states. They denoted this feature as the large $c$ ambiguity in identity Virasoro block. Their partial resolution was that at finite $c$, vacuum Virasoro blocks do not have any periodic singularities and HHLL should not decay\footnote{Any perturbative correction to identity blocks \cite{Fitzpatrick:2015dlt} or any correction coming from non-identity block would not change the decaying nature, see for instance \cite{Pal:2022uoo}.}. In other way, this form of typicality is a large $c$ artifact. Those pure states are dual to (planar) BTZ and they are candidates of typical microstates having a horizon.

Here, we would like to confront these puzzling situations from modular quantization. In the regulated Hilbert space $\mathcal{H}_{\epsilon}$, the (vacuum) descendants are typical states in $\epsilon \rightarrow 0$ limit as we have shown. Even all other primary states $|h,h\rangle^{\epsilon}$ \footnote{These are global descendants in radial quantization as explained in appendix.} with $h<c/24$ behaves as thermal states in the $\epsilon \rightarrow 0$ limit. This explains the first puzzle. However the second puzzle for $h>c/24$ is still there in modular quantization. As we obtained the two point function in heavy ($h>c/24$) primaries will not decay and they do not have periodic singularity in Euclidean time at large $c$ as opposed to radial quantization. In other words, the typical state in radial quantization behaves like a scar in the modular quantization in semiclassical limit. We may justify this along the line of \cite{Fitzpatrick:2016ive} by denoting this as an artifact of large $c$ vacuum Virasoro block. In fact, in modular quantization the periodic behavior of Euclidean time should be the generic feature in all pure state correlatiors. We again expect at finite $c$, the HHLL correlator will behave thermally. From this observations, we would want to emphasize the following (re)statements:

\begin{itemize}
    \item The high energy primaries in radial quantization are not generic typical state in the sense of local ETH\cite{Lashkari:2016vgj}. The typicality is an artifact of large $c$ behavior and it is a frame dependent statement. The large $c$ typicality in radial quantization becomes a large $c$ scar in modular quantization. The horizon or conical defect in the dual description of those states becomes observer dependent statement.

    \item Modular Hamiltonian could be an example of chaotic Hamiltonian. Study of chaotic properties in the eigenstates of $H_{\epsilon}$ would be an interesting future study. In the dual stretched horizon background, some chaotic properties have already been observed recently \cite{Das:2022evy, Das:2023ulz, Das:2023xjr}\footnote{See also the recent paper \cite{Das:2024fwg} for a generalization of similar story in higher dimensions.}. The scrambling in modular time \cite{Das:2022jrr, Khetrapal:2022dzy} as well as Krylov complexity for such Hamiltonian dynamics \cite{Malvimat:2024vhr} has also been explored.
\end{itemize}

\item \underline{\textit{Black hole microstates with no horizon:}} From all of our results and discussions, we should think of the planar BTZ as the TFD state in canonical ensemble \cite{Das:2024vqe}, which coincides with the vacuum in $\epsilon \rightarrow 0$ limit. On the other hand, we must think BTZ as the TFD representation of microcanonical ensemble at some typical energy $E(\sim c/12)$ \cite{Chandrasekaran:2022eqq}:
\begin{align}
    |0\rangle_{BTZ} = e^{-S(E)}\sum_{i}e^{-\beta(E_{i}-E)/2}f(E_{i}-E)|E_{i}\rangle_{L}^{\epsilon}|E_{i}\rangle_{R}^{\epsilon}
\end{align}
Here $f$ is a smooth and invertible function satisfying $\int dx|f(x)|^{2} =1$. $L$ and $R$ denotes left and right Hilbert space at finite $\epsilon$. In contrast to canonical picture(or AdS-Rindler), here $\mathcal{H}_{\epsilon} = \mathcal{H}_{L}^{\epsilon}\otimes \mathcal{H}_{R}^{\epsilon}$\footnote{Note that, the actual description of the Hilbert space should be given by direct sum over boundary conditions\cite{Das:2024vqe}. However, here we are considering only one particular boundary condition that determines the limit $|0_{\epsilon}\rangle \rightarrow |0\rangle$ in $\epsilon \rightarrow 0$.}. The states $|E_{i}\rangle^{\epsilon}$ consists of (vacuum) descendants with energy $c/12$ and primaries $|h,h\rangle^{\epsilon}$ with $h>\frac{c}{24}$. In $\epsilon \rightarrow 0$ the (vacuum) descendants becomes typical microstates for all $c$, whereas the other primaries will not be typical even in $c \rightarrow \infty$. We have conjectured the typical microstates can be described as fluctuation of free scalars in (planar) BTZ stretched horizon background. However, both of these kind of microstates may not have semiclassical description in terms of capped off geometry in Einstein gravity regime. The primary reason behind this is the fact that these states belong to only in the bigger regulated Hilbert space $\mathcal{H}_{\epsilon}$. In strict $\epsilon \rightarrow 0$, these states are getting mapped to states with known semiclassical dual as we discussed in earlier section. Hence the finite $\epsilon$ states do not belong to unregulated Hilbert space $\mathcal{H}_{\epsilon\rightarrow 0}$ and hence should not be described by weak gravity approximation in the bulk. This seems to be puzzling in the context of AdS/CFT. However there could be two possibilities to overcome this apparent puzzle. One way to fix this is by demanding $\epsilon \rightarrow 0$ limit to be proportional to $c \rightarrow \infty$. However, from the discussion of microcanonical entropy, we can simply discard this for BTZ, since we argued $\epsilon$ to be fixed at $\mathcal{O}(1)$. The another possibility is to declare that there will be no semiclassical description of finite $\epsilon$ states and taking $c \rightarrow \infty$ with finite $\epsilon$ is not meaningful and physical. Our discussion indicates that this could be the most possible answer to the puzzle. We admit that only some string theoretic top-down realization of AdS/CFT for such regulated Hilbert space would shed light to this. We believe, our construction of such microstates in the regulated Hilbert space might have a connection to non-supersymmetric version of Fuzzball solution\cite{Lunin:2001jy}. However, the previous discussion discards the possibility\cite{Raju:2018xue} of such `fuzzball' solution is accessible in Einstein gravity regime \cite{Bena:2013dka}. From our analysis, we also expect the BTZ can also be realized in similar type of alternative quantization where the spatial direction ($s$) of the quantization contour will be compactified. That would shed more light on Maldacena's information paradox\cite{Maldacena:2001kr}. It would be great to understand the emergence of Planckian stretched horizon as used in our previous analysis \cite{Burman:2023kko}, which is important to understand the emergence of type III$_{1}$ factor in BTZ(where unlike planar BTZ, the boundary CFT type III$_{1}$ factor is also emergent only in $c \rightarrow \infty$).
 
 \end{itemize}

\section{Acknowledgement}
The author(SD) of this paper strongly believes that any scientific progress is a collective and collaborative process where the authorship is an effort to consolidate all the ideas in a single manuscript. In this regard, SD would like to acknowledge people who came across at different stages with ideas, discussions that helped us to write this paper. SD would like to thank Vaibhav Burman, Bobby Ezhuthachan, Chethan Krishnan, Somnath Porey and Baishali Roy for many important discussions and collaboration in the course of this project. SD is grateful to Vaibhav for checking several calculations and finding some typos in this draft. Especially, SD would like to thank Bobby for many related discussion over years, since the idea of thermal quantization in CFT was always there which we discussed a lot. SD would like to acknowledge the physics department of RKMVERI for a visit, where the project was discussed at the primitive stage. SD would also like to thank Diptarka Das, Justin David, Anirban Dinda, Sanchari Pal, Partha Paul and Krishnendu Sengupta for many related discussions that help us to write this paper. SD would also like to thank Diptarka for a visit to IIT Kanpur where part of the work has been done. SD would like to acknowledge the organisers of `Workshop on observable algebras in field theory and gravity' for the stay at IIT Mandi where part of the work has been done. The research work by SD is supported by a DST INSPIRE Faculty fellowship.
\appendix
\section{The other vacuum of the quantization}\label{app1}
Here we will show explicitly that $|h,h\rangle$ which is defined in (\ref{in})(in the $\epsilon \rightarrow 0$ limit) indeed is the vacuum as we observed in (\ref{new vac}). This in principle provides a check of the correctness of the method of quantization we proposed.

Let us begin with an ansatz of a new vacuum state of the following form:
\begin{align}
    |0\rangle_{new} = \sum_{n,m}a_{n}b_{m} L_{-1}^{n} \Bar{L}_{-1}^{m}\mathcal{O}_{h,h}(0,0)|0\rangle
\end{align}
such that it satisfies
\begin{align}\label{demand vac}
    H|0\rangle_{new} = 0
\end{align}
where $H$ is defined as in (\ref{Ham}). Since the Hamiltonian is a sum of holomorphic and antiholomorphic sector, the choice of such a state is natural and well-motivated. The holomorphic sector of the Hamiltonian acts as
\begin{align}
    &(\alpha L_{0}+ \beta(L_{1}+L_{-1})) \sum_{n}a_{n}L^{n}_{-1} \mathcal{O}_{h,h}(0,0)|0\rangle \nonumber \\
    & = \sum_{n} \Big[(\alpha n a_{n}+\alpha a_{n} h)L^{n}_{-1} + (\beta a_{n} n(n-1)+2\beta n a_{n} h)L^{n-1}_{-1} + \beta a_{n}L^{n+1}_{-1}\Big]\mathcal{O}_{h,h}(0,0)|0\rangle \nonumber \\
    & = \sum_{n}\Big[\alpha n a_{n} + \alpha a_{n} h+ \beta a_{n+1} n(n+1) + 2\beta a_{n+1} (n+1)h+\beta a_{n-1}\Big] L^{n}_{-1}\mathcal{O}_{h,h}(0,0)|0\rangle 
\end{align}
Here in the second line we used the following commutator relations of global Virasoro generators
\begin{align}
    &[L_{1},L^{n}_{-1}] = n(n-1)L^{n-1}_{-1} + 2nL^{n-1}_{-1}L_{0} \nonumber \\
    & [L_{0},L_{-1}^{n}] = n L^{n}_{-1}
\end{align}
Similarly we could also obtain the action on the state by the anti-holomorphioc sector of the Hamiltonian. After summing both we get
\begin{align}
   & H|0\rangle_{new} \nonumber \\
   &= \sum_{n,m}\Big[b_{m}\Big( a_{n+1}(\beta n(n+1) +2\beta(n+1)h)+a_{n}(\alpha n +\alpha h) +a_{n-1}\beta\Big) + \nonumber \\ &+a_{n}\Big( b_{m+1}(\beta m(m+1) +2\beta(m+1)h)+b_{m}(\alpha m +\alpha h) +b_{m-1}\beta\Big)\Big]L^{n}_{-1}\bar{L}_{-1}^{m}\mathcal{O}_{h,h}(0,0)|0\rangle
\end{align}
If we choose $a_{n} = \frac{a^{n}}{n!}$ and $b_{m} = \frac{b^{m}}{m!}$, we will have
\begin{align}
  & H|0\rangle_{new} \nonumber \\
   &=   \sum_{n,m} \frac{b^{m}a^{n}}{m!n!}\Big( 2\beta a h+\alpha h +n(a\beta + \alpha + \frac{\beta}{a}) +  2\beta b h+\alpha h +m(b\beta + \alpha + \frac{\beta}{b})\Big)L^{n}_{-1}\bar{L}_{-1}^{m}\mathcal{O}_{h,h}(0,0)|0\rangle
\end{align}
One would readily see that there exists one unique solution to (\ref{demand vac}) which is
\begin{align}
 \alpha = \bar{\beta} = \frac{-\alpha + \sqrt{\alpha^{2}-4\beta^{2}}}{2\beta}  = z_{+} 
\end{align}
This gives the final form of the desired state as
\begin{align}
    |0\rangle_{new} = \sum_{n,m} \frac{z_{+}^{n}}{n!}\frac{z_{-}^{m}}{m!}L^{n}_{-1}\bar{L}_{-1}^{m}\mathcal{O}_{h,h}(0,0)|0\rangle =|h,h\rangle^{\epsilon \rightarrow 0}
\end{align}

One can also check that $[H,\mathcal{O}_{h,h}(z_{+},z_{-})]$ is vanishing itself. Thus these operators are zero modes or symmetry generators of the Hamiltonian. These operators can be thought of conserved charges which do not decay in time generically\cite{Lashkari:2019ixo}. However in our present work, we do not consider such operators to be a part of the algebra rather they create some cyclic and separating states after acting on the vacuum. For certain $h\sim \mathcal{O}(c)$, those states can be thought of as a new vacuum of the modular Hamiltonian $H$ in semiclassical limit as they decay in time and behaves similarly to a thermal correlator.


\begin{thebibliography}{99}

\bibitem{Maldacena:1997re}
J.~M.~Maldacena,
``The Large N limit of superconformal field theories and supergravity,''
Adv. Theor. Math. Phys. \textbf{2} (1998), 231-252
doi:10.4310/ATMP.1998.v2.n2.a1
[arXiv:hep-th/9711200 [hep-th]].

\bibitem{Aharony:1999ti}
O.~Aharony, S.~S.~Gubser, J.~M.~Maldacena, H.~Ooguri and Y.~Oz,
``Large N field theories, string theory and gravity,''
Phys. Rept. \textbf{323} (2000), 183-386
doi:10.1016/S0370-1573(99)00083-6
[arXiv:hep-th/9905111 [hep-th]].

\bibitem{Heemskerk:2009pn}
I.~Heemskerk, J.~Penedones, J.~Polchinski and J.~Sully,
``Holography from Conformal Field Theory,''
JHEP \textbf{10} (2009), 079
doi:10.1088/1126-6708/2009/10/079
[arXiv:0907.0151 [hep-th]].

\bibitem{El-Showk:2011yvt}
S.~El-Showk and K.~Papadodimas,
``Emergent Spacetime and Holographic CFTs,''
JHEP \textbf{10} (2012), 106
doi:10.1007/JHEP10(2012)106
[arXiv:1101.4163 [hep-th]].

\bibitem{Maldacena:2001kr}
J.~M.~Maldacena,
``Eternal black holes in anti-de Sitter,''
JHEP \textbf{04} (2003), 021
doi:10.1088/1126-6708/2003/04/021
[arXiv:hep-th/0106112 [hep-th]].

\bibitem{Brown:1986nw}
J.~D.~Brown and M.~Henneaux,
``Central Charges in the Canonical Realization of Asymptotic Symmetries: An Example from Three-Dimensional Gravity,''
Commun. Math. Phys. \textbf{104} (1986), 207-226
doi:10.1007/BF01211590

\bibitem{Banados:1994tn}
M.~Banados,
``Global charges in Chern-Simons field theory and the (2+1) black hole,''
Phys. Rev. D \textbf{52} (1996), 5816-5825
doi:10.1103/PhysRevD.52.5816

\bibitem{Banados:1992gq}
M.~Banados, M.~Henneaux, C.~Teitelboim and J.~Zanelli,
``Geometry of the (2+1) black hole,''
Phys. Rev. D \textbf{48} (1993), 1506-1525
[erratum: Phys. Rev. D \textbf{88} (2013), 069902]
doi:10.1103/PhysRevD.48.1506
[arXiv:gr-qc/9302012 [gr-qc]].

\bibitem{Fitzpatrick:2014vua}
A.~L.~Fitzpatrick, J.~Kaplan and M.~T.~Walters,
``Universality of Long-Distance AdS Physics from the CFT Bootstrap,''
JHEP \textbf{08} (2014), 145
doi:10.1007/JHEP08(2014)145
[arXiv:1403.6829 [hep-th]].

\bibitem{Fitzpatrick:2015zha}
A.~L.~Fitzpatrick, J.~Kaplan and M.~T.~Walters,
``Virasoro Conformal Blocks and Thermality from Classical Background Fields,''
JHEP \textbf{11} (2015), 200
doi:10.1007/JHEP11(2015)200
[arXiv:1501.05315 [hep-th]].

\bibitem{Cardy:1986ie}
J.~L.~Cardy,
``Operator Content of Two-Dimensional Conformally Invariant Theories,''
Nucl. Phys. B \textbf{270} (1986), 186-204
doi:10.1016/0550-3213(86)90552-3


\bibitem{Leutheusser:2021qhd}
S.~Leutheusser and H.~Liu,
``Causal connectability between quantum systems and the black hole interior in holographic duality,''
Phys. Rev. D \textbf{108} (2023) no.8, 086019
doi:10.1103/PhysRevD.108.086019
[arXiv:2110.05497 [hep-th]].

\bibitem{Leutheusser:2021frk}
S.~A.~W.~Leutheusser,
``Emergent Times in Holographic Duality,''
Phys. Rev. D \textbf{108} (2023) no.8, 086020
doi:10.1103/PhysRevD.108.086020
[arXiv:2112.12156 [hep-th]].


\bibitem{Witten:2018zxz}
E.~Witten,
``APS Medal for Exceptional Achievement in Research: Invited article on entanglement properties of quantum field theory,''
Rev. Mod. Phys. \textbf{90} (2018) no.4, 045003
doi:10.1103/RevModPhys.90.045003
[arXiv:1803.04993 [hep-th]].

\bibitem{Papadodimas:2012aq}
K.~Papadodimas and S.~Raju,
``An Infalling Observer in AdS/CFT,''
JHEP \textbf{10} (2013), 212
doi:10.1007/JHEP10(2013)212
[arXiv:1211.6767 [hep-th]].

\bibitem{Furuya:2023fei}
K.~Furuya, N.~Lashkari, M.~Moosa and S.~Ouseph,
``Information loss, mixing and emergent type III$_{1}$ factors,''
JHEP \textbf{08} (2023), 111
doi:10.1007/JHEP08(2023)111
[arXiv:2305.16028 [hep-th]].

\bibitem{Burman:2023kko}
V.~Burman, S.~Das and C.~Krishnan,
``A smooth horizon without a smooth horizon,''
JHEP \textbf{03} (2024), 014
doi:10.1007/JHEP03(2024)014
[arXiv:2312.14108 [hep-th]].

\bibitem{Banerjee:2024dpl}
S.~Banerjee, S.~Das, M.~Dorband and A.~Kundu,
``Brickwall, Normal Modes and Emerging Thermality,''
[arXiv:2401.01417 [hep-th]].

\bibitem{tHooft:1984kcu}
G.~'t Hooft,
``On the Quantum Structure of a Black Hole,''
Nucl. Phys. B \textbf{256} (1985), 727-745
doi:10.1016/0550-3213(85)90418-3


\bibitem{Solodukhin:2005qy}
S.~N.~Solodukhin,
``Restoring unitarity in BTZ black hole,''
Phys. Rev. D \textbf{71} (2005), 064006
doi:10.1103/PhysRevD.71.064006
[arXiv:hep-th/0501053 [hep-th]].

\bibitem{Krishnan:2023jqn}
C.~Krishnan and P.~S.~Pathak,
``Normal modes of the stretched horizon: a bulk mechanism for black hole microstate level spacing,''
JHEP \textbf{03} (2024), 162
doi:10.1007/JHEP03(2024)162
[arXiv:2312.14109 [hep-th]].

\bibitem{Mukohyama:1998rf}
S.~Mukohyama and W.~Israel,
``Black holes, brick walls and the Boulware state,''
Phys. Rev. D \textbf{58} (1998), 104005
doi:10.1103/PhysRevD.58.104005
[arXiv:gr-qc/9806012 [gr-qc]].

\bibitem{Burman:2024egy}
V.~Burman and C.~Krishnan,
``A Bottom-Up Approach to Black Hole Microstates,''
[arXiv:2409.05850 [hep-th]].

\bibitem{Calabrese:2016xau}
P.~Calabrese and J.~Cardy,
``Quantum quenches in 1  +  1 dimensional conformal field theories,''
J. Stat. Mech. \textbf{1606} (2016) no.6, 064003
doi:10.1088/1742-5468/2016/06/064003
[arXiv:1603.02889 [cond-mat.stat-mech]].


\bibitem{Hartman:2013qma}
T.~Hartman and J.~Maldacena,
``Time Evolution of Entanglement Entropy from Black Hole Interiors,''
JHEP \textbf{05} (2013), 014
doi:10.1007/JHEP05(2013)014
[arXiv:1303.1080 [hep-th]].

\bibitem{Almheiri:2018ijj}
A.~Almheiri, A.~Mousatov and M.~Shyani,
``Escaping the interiors of pure boundary-state black holes,''
JHEP \textbf{02} (2023), 024
doi:10.1007/JHEP02(2023)024
[arXiv:1803.04434 [hep-th]].

\bibitem{Tada:2019rls}
T.~Tada,
``Time development of conformal field theories associated with $L_{1}$ and $L_{-1}$ operators,''
J. Phys. A \textbf{53} (2020) no.25, 255401
doi:10.1088/1751-8121/ab8c63
[arXiv:1904.12414 [hep-th]].

\bibitem{Agia:2022srj}
N.~Agia and D.~L.~Jafferis,
``Angular Quantization in CFT,''
[arXiv:2204.11872 [hep-th]].

\bibitem{Agia:2024wxx}
N.~Agia and D.~L.~Jafferis,
``The 2d Free Boson Minkowski CFT with Asymptotic Charges,''
[arXiv:2402.05167 [hep-th]].

\bibitem{Cardy:2016fqc}
J.~Cardy and E.~Tonni,
``Entanglement hamiltonians in two-dimensional conformal field theory,''
J. Stat. Mech. \textbf{1612} (2016) no.12, 123103
doi:10.1088/1742-5468/2016/12/123103
[arXiv:1608.01283 [cond-mat.stat-mech]].

\bibitem{deBoer:2021zlm}
J.~de Boer, R.~Esp\'\i{}ndola, B.~Najian, D.~Patramanis, J.~van der Heijden and C.~Zukowski,
``Virasoro entanglement Berry phases,''
JHEP \textbf{03} (2022), 179
doi:10.1007/JHEP03(2022)179
[arXiv:2111.05345 [hep-th]].

\bibitem{Das:2024vqe}
S.~Das, B.~Ezhuthachan, S.~Porey and B.~Roy,
``Notes on heating phase dynamics in Floquet CFTs and Modular quantization,''
[arXiv:2406.10899 [hep-th]].

\bibitem{Sorce:2023gio}
J.~Sorce,
``An intuitive construction of modular flow,''
JHEP \textbf{12} (2023), 079
doi:10.1007/JHEP12(2023)079
[arXiv:2309.16766 [hep-th]].

\bibitem{Belavin:1984vu}
A.~A.~Belavin, A.~M.~Polyakov and A.~B.~Zamolodchikov,
``Infinite Conformal Symmetry in Two-Dimensional Quantum Field Theory,''
Nucl. Phys. B \textbf{241} (1984), 333-380
doi:10.1016/0550-3213(84)90052-X

\bibitem{Okunishi:2016zat}
K.~Okunishi,
``Sine-square deformation and M\"obius quantization of 2D conformal field theory,''
PTEP \textbf{2016} (2016) no.6, 063A02
doi:10.1093/ptep/ptw060
[arXiv:1603.09543 [hep-th]].

\bibitem{Ishibashi:2015jba}
N.~Ishibashi and T.~Tada,
``Infinite circumference limit of conformal field theory,''
J. Phys. A \textbf{48} (2015) no.31, 315402
doi:10.1088/1751-8113/48/31/315402
[arXiv:1504.00138 [hep-th]].

\bibitem{Ishibashi:2016bey}
N.~Ishibashi and T.~Tada,
``Dipolar quantization and the infinite circumference limit of two-dimensional conformal field theories,''
Int. J. Mod. Phys. A \textbf{31} (2016) no.32, 1650170
doi:10.1142/S0217751X16501700
[arXiv:1602.01190 [hep-th]].

\bibitem{Das:2020goe}
S.~Das, B.~Ezhuthachan, S.~Porey and B.~Roy,
``Virasoro algebras, kinematic space and the spectrum of modular Hamiltonians in CFT$_{2}$,''
JHEP \textbf{21} (2020), 123
doi:10.1007/JHEP08(2021)123
[arXiv:2101.10211 [hep-th]].

\bibitem{Aharony:2003sx}
O.~Aharony, J.~Marsano, S.~Minwalla, K.~Papadodimas and M.~Van Raamsdonk,
``The Hagedorn - deconfinement phase transition in weakly coupled large N gauge theories,''
Adv. Theor. Math. Phys. \textbf{8} (2004), 603-696
doi:10.4310/ATMP.2004.v8.n4.a1
[arXiv:hep-th/0310285 [hep-th]].

\bibitem{Ohmori:2014eia}
K.~Ohmori and Y.~Tachikawa,
``Physics at the entangling surface,''
J. Stat. Mech. \textbf{1504} (2015), P04010
doi:10.1088/1742-5468/2015/04/P04010
[arXiv:1406.4167 [hep-th]].

\bibitem{Czech:2017zfq}
B.~Czech, L.~Lamprou, S.~Mccandlish and J.~Sully,
``Modular Berry Connection for Entangled Subregions in AdS/CFT,''
Phys. Rev. Lett. \textbf{120} (2018) no.9, 091601
doi:10.1103/PhysRevLett.120.091601
[arXiv:1712.07123 [hep-th]].

\bibitem{Das:2019iit}
S.~Das and B.~Ezhuthachan,
``Spectrum of Modular Hamiltonian in the Vacuum and Excited States,''
JHEP \textbf{10} (2019), 009
doi:10.1007/JHEP10(2019)009
[arXiv:1906.00726 [hep-th]].

\bibitem{Ishibashi:1988kg}
N.~Ishibashi,
``The Boundary and Crosscap States in Conformal Field Theories,''
Mod. Phys. Lett. A \textbf{4} (1989), 251
doi:10.1142/S0217732389000320

\bibitem{Castro:2018srf}
A.~Castro, N.~Iqbal and E.~Llabr\'es,
``Wilson lines and Ishibashi states in AdS$_{3}$/CFT$_{2}$,''
JHEP \textbf{09} (2018), 066
doi:10.1007/JHEP09(2018)066
[arXiv:1805.05398 [hep-th]].

\bibitem{Strominger:1997eq}
A.~Strominger,
``Black hole entropy from near horizon microstates,''
JHEP \textbf{02} (1998), 009
doi:10.1088/1126-6708/1998/02/009
[arXiv:hep-th/9712251 [hep-th]].

\bibitem{Das:2022pez}
S.~Das, B.~Ezhuthachan, A.~Kundu, S.~Porey, B.~Roy and K.~Sengupta,
``Brane detectors of a dynamical phase transition in a driven CFT,''
SciPost Phys. \textbf{15} (2023) no.5, 202
doi:10.21468/SciPostPhys.15.5.202
[arXiv:2212.04201 [hep-th]].

\bibitem{Roberts:2012aq}
M.~M.~Roberts,
``Time evolution of entanglement entropy from a pulse,''
JHEP \textbf{12} (2012), 027
doi:10.1007/JHEP12(2012)027
[arXiv:1204.1982 [hep-th]].

\bibitem{Asplund:2014coa}
C.~T.~Asplund, A.~Bernamonti, F.~Galli and T.~Hartman,
``Holographic Entanglement Entropy from 2d CFT: Heavy States and Local Quenches,''
JHEP \textbf{02} (2015), 171
doi:10.1007/JHEP02(2015)171
[arXiv:1410.1392 [hep-th]].

\bibitem{Soni:2023fke}
R.~M.~Soni,
``A type I approximation of the crossed product,''
JHEP \textbf{01} (2024), 123
doi:10.1007/JHEP01(2024)123
[arXiv:2307.12481 [hep-th]].

\bibitem{Casini:2011kv}
H.~Casini, M.~Huerta and R.~C.~Myers,
``Towards a derivation of holographic entanglement entropy,''
JHEP \textbf{05} (2011), 036
doi:10.1007/JHEP05(2011)036
[arXiv:1102.0440 [hep-th]].

\bibitem{Czech:2012be}
B.~Czech, J.~L.~Karczmarek, F.~Nogueira and M.~Van Raamsdonk,
``Rindler Quantum Gravity,''
Class. Quant. Grav. \textbf{29} (2012), 235025
doi:10.1088/0264-9381/29/23/235025
[arXiv:1206.1323 [hep-th]].

\bibitem{Ryu:2006bv}
S.~Ryu and T.~Takayanagi,
``Holographic derivation of entanglement entropy from AdS/CFT,''
Phys. Rev. Lett. \textbf{96} (2006), 181602
doi:10.1103/PhysRevLett.96.181602
[arXiv:hep-th/0603001 [hep-th]].

\bibitem{Hubeny:2007xt}
V.~E.~Hubeny, M.~Rangamani and T.~Takayanagi,
``A Covariant holographic entanglement entropy proposal,''
JHEP \textbf{07} (2007), 062
doi:10.1088/1126-6708/2007/07/062
[arXiv:0705.0016 [hep-th]].

\bibitem{Susskind:1994sm}
L.~Susskind and J.~Uglum,
``Black hole entropy in canonical quantum gravity and superstring theory,''
Phys. Rev. D \textbf{50} (1994), 2700-2711
doi:10.1103/PhysRevD.50.2700
[arXiv:hep-th/9401070 [hep-th]].

\bibitem{Bahiru:2022mwh}
E.~Bahiru,
``Algebra of operators in an AdS-Rindler wedge,''
JHEP \textbf{06} (2023), 197
doi:10.1007/JHEP06(2023)197
[arXiv:2208.04258 [hep-th]].

\bibitem{Leutheusser:2022bgi}
S.~Leutheusser and H.~Liu,
``Subalgebra-subregion duality: emergence of space and time in holography,''
[arXiv:2212.13266 [hep-th]].

\bibitem{Jensen:2023yxy}
K.~Jensen, J.~Sorce and A.~J.~Speranza,
``Generalized entropy for general subregions in quantum gravity,''
JHEP \textbf{12} (2023), 020
doi:10.1007/JHEP12(2023)020
[arXiv:2306.01837 [hep-th]].

\bibitem{AliAhmad:2023etg}
S.~Ali Ahmad and R.~Jefferson,
``Crossed product algebras and generalized entropy for subregions,''
SciPost Phys. Core \textbf{7} (2024), 020
doi:10.21468/SciPostPhysCore.7.2.020
[arXiv:2306.07323 [hep-th]].

\bibitem{Witten:2021unn}
E.~Witten,
``Gravity and the crossed product,''
JHEP \textbf{10} (2022), 008
doi:10.1007/JHEP10(2022)008
[arXiv:2112.12828 [hep-th]].

\bibitem{Chandrasekaran:2022eqq}
V.~Chandrasekaran, G.~Penington and E.~Witten,
``Large N algebras and generalized entropy,''
JHEP \textbf{04} (2023), 009
doi:10.1007/JHEP04(2023)009
[arXiv:2209.10454 [hep-th]].

\bibitem{Kudler-Flam:2023qfl}
J.~Kudler-Flam, S.~Leutheusser and G.~Satishchandran,
``Generalized Black Hole Entropy is von Neumann Entropy,''
[arXiv:2309.15897 [hep-th]].

\bibitem{Festuccia:2005pi}
G.~Festuccia and H.~Liu,
``Excursions beyond the horizon: Black hole singularities in Yang-Mills theories. I.,''
JHEP \textbf{04} (2006), 044
doi:10.1088/1126-6708/2006/04/044
[arXiv:hep-th/0506202 [hep-th]].

\bibitem{Festuccia:2006sa}
G.~Festuccia and H.~Liu,
``The Arrow of time, black holes, and quantum mixing of large N Yang-Mills theories,''
JHEP \textbf{12} (2007), 027
doi:10.1088/1126-6708/2007/12/027
[arXiv:hep-th/0611098 [hep-th]].


\bibitem{Brehm:2018ipf}
E.~M.~Brehm, D.~Das and S.~Datta,
``Probing thermality beyond the diagonal,''
Phys. Rev. D \textbf{98} (2018) no.12, 126015
doi:10.1103/PhysRevD.98.126015
[arXiv:1804.07924 [hep-th]].

\bibitem{Basu:2017kzo}
P.~Basu, D.~Das, S.~Datta and S.~Pal,
``Thermality of eigenstates in conformal field theories,''
Phys. Rev. E \textbf{96} (2017) no.2, 022149
doi:10.1103/PhysRevE.96.022149
[arXiv:1705.03001 [hep-th]].

\bibitem{Datta:2019jeo}
S.~Datta, P.~Kraus and B.~Michel,
``Typicality and thermality in 2d CFT,''
JHEP \textbf{07} (2019), 143
doi:10.1007/JHEP07(2019)143
[arXiv:1904.00668 [hep-th]].

\bibitem{Fitzpatrick:2016ive}
A.~L.~Fitzpatrick, J.~Kaplan, D.~Li and J.~Wang,
``On information loss in AdS$_{3}$/CFT$_{2}$,''
JHEP \textbf{05} (2016), 109
doi:10.1007/JHEP05(2016)109
[arXiv:1603.08925 [hep-th]].

\bibitem{Fitzpatrick:2015dlt}
A.~L.~Fitzpatrick and J.~Kaplan,
``Conformal Blocks Beyond the Semi-Classical Limit,''
JHEP \textbf{05} (2016), 075
doi:10.1007/JHEP05(2016)075
[arXiv:1512.03052 [hep-th]].

\bibitem{Pal:2022uoo}
S.~Pal,
``Finite temperature corrections to black hole quasinormal modes from 2D CFT,''
JHEP \textbf{08} (2022), 150
doi:10.1007/JHEP08(2022)150
[arXiv:2201.10264 [hep-th]].

\bibitem{Lashkari:2016vgj}
N.~Lashkari, A.~Dymarsky and H.~Liu,
``Eigenstate Thermalization Hypothesis in Conformal Field Theory,''
J. Stat. Mech. \textbf{1803} (2018) no.3, 033101
doi:10.1088/1742-5468/aab020
[arXiv:1610.00302 [hep-th]].

\bibitem{Das:2022evy}
S.~Das, C.~Krishnan, A.~P.~Kumar and A.~Kundu,
``Synthetic fuzzballs: a linear ramp from black hole normal modes,''
JHEP \textbf{01} (2023), 153
doi:10.1007/JHEP01(2023)153
[arXiv:2208.14744 [hep-th]].

\bibitem{Das:2023ulz}
S.~Das, S.~K.~Garg, C.~Krishnan and A.~Kundu,
``Fuzzballs and random matrices,''
JHEP \textbf{10} (2023), 031
doi:10.1007/JHEP10(2023)031
[arXiv:2301.11780 [hep-th]].

\bibitem{Das:2023xjr}
S.~Das and A.~Kundu,
``Brickwall in rotating BTZ: a dip-ramp-plateau story,''
JHEP \textbf{02} (2024), 049
doi:10.1007/JHEP02(2024)049
[arXiv:2310.06438 [hep-th]].

\bibitem{Das:2024fwg}
S.~Das, S.~Porey and B.~Roy,
``Brick Wall in AdS-Schwarzschild Black Hole: Normal Modes and Emerging Thermality,''
[arXiv:2409.05519 [hep-th]].

\bibitem{Das:2022jrr}
S.~Das, B.~Ezhuthachan, A.~Kundu, S.~Porey, B.~Roy and K.~Sengupta,
``Out-of-Time-Order correlators in driven conformal field theories,''
JHEP \textbf{08} (2022), 221
doi:10.1007/JHEP08(2022)221
[arXiv:2202.12815 [hep-th]].


\bibitem{Khetrapal:2022dzy}
S.~Khetrapal,
``Chaos and operator growth in 2d CFT,''
JHEP \textbf{03} (2023), 176
doi:10.1007/JHEP03(2023)176
[arXiv:2210.15860 [hep-th]].

\bibitem{Malvimat:2024vhr}
V.~Malvimat, S.~Porey and B.~Roy,
``Krylov Complexity in $2d$ CFTs with SL$(2,\mathbb{R})$ deformed Hamiltonians,''
[arXiv:2402.15835 [hep-th]].

\bibitem{Lunin:2001jy}
O.~Lunin and S.~D.~Mathur,
``AdS / CFT duality and the black hole information paradox,''
Nucl. Phys. B \textbf{623} (2002), 342-394
doi:10.1016/S0550-3213(01)00620-4
[arXiv:hep-th/0109154 [hep-th]].

\bibitem{Raju:2018xue}
S.~Raju and P.~Shrivastava,
``Critique of the fuzzball program,''
Phys. Rev. D \textbf{99} (2019) no.6, 066009
doi:10.1103/PhysRevD.99.066009
[arXiv:1804.10616 [hep-th]].


\bibitem{Bena:2013dka}
I.~Bena and N.~P.~Warner,
``Resolving the Structure of Black Holes: Philosophizing with a Hammer,''
[arXiv:1311.4538 [hep-th]].


\bibitem{Lashkari:2019ixo}
N.~Lashkari,
``Modular zero modes and sewing the states of QFT,''
JHEP \textbf{21} (2020), 189
doi:10.1007/JHEP04(2021)189
[arXiv:1911.11153 [hep-th]].

\end{thebibliography}
\end{document}